\documentclass[journal]{IEEEtran}
\usepackage{graphicx}
\usepackage{amsmath}
\usepackage{amssymb}
\usepackage{url}
\usepackage[noadjust]{cite}

\usepackage{array} 
\usepackage{algorithm,algpseudocode}
\newtheorem{theorem}{Theorem}[section]

\newtheorem{lemma}[theorem]{Lemma}
\newtheorem{proof}[theorem]{Proof}

\ifCLASSINFOpdf
\else
\fi
\begin{document}
%
\title{Distributed Obstacle and Multi-Robot Collision Avoidance in Uncertain Environments}
%
%
%

\author{Vu Phi Tran,~Matthew~Garratt and~Ian~R.Petersen%
	\thanks{Vu~Phi~Tran and~Matthew~Garratt are with the School of Engineering and Information Technology, University of New South Wales, Australia and~Ian~R.Petersen is with the Research School of Engineering, Australian National University, Australia. {Phi.Tran}@student.adfa.edu.au, {M.Garratt}@adfa.edu.au and {I.R.Petersen}@gmail.com}
}

%
%

\markboth{}%
{Shell \MakeLowercase{\textit{et al.}}: Bare Demo of IEEEtran.cls for IEEE Journals}
%



\maketitle
\begin{abstract}
This paper tackles the distributed leader-follower (L-F) control problem for heterogeneous mobile robots in unknown environments requiring obstacle avoidance, inter-robot collision avoidance, and reliable robot communications. To prevent an inter-robot collision, we employ a virtual propulsive force between robots. For obstacle avoidance, we present a novel distributed Negative-Imaginary (NI) variant formation tracking control approach and a dynamic network topology methodology which allows the formation to change its shape and the robot to switch their roles. In the case of communication or sensor loss, a UAV, controlled by a Strictly-Negative-Imaginary (SNI) controller with good wind resistance characteristics, is utilized to track the position of the UGV formation using its camera. Simulations and indoor experiments have been conducted to validate the proposed methods.
\end{abstract}


%
\IEEEpeerreviewmaketitle

\section{Introduction}

\IEEEPARstart{M}{ulti-robot} systems are one of crucial requirements for many applications such as surveillance, target tracking, self-localization and mapping (SLAM), and disaster response. Formation keeping, which is one of the most common approaches for multi-robot control, has been studied extensively in recent years \cite{Rahimi14, Zhao17, Dong16, Dong17}. It is pointed out in \cite{Beard01} that there are three fundamental formation control strategies which are commonly used for a multi-vehicle system: virtual leader, virtual structure and behavior, and leader-follower. Because of its simplicity and effectiveness in practical applications, the leader-follower (L-F) approach is preferred by researchers. As a result, a large body of work related to L-F applications has emerged, ranging from searching, surveillance, inspection, and exploration \cite{Panagou12}. According to \cite{Ni10, Hou15}, the L-F consensus problem can be summarized as follows: all followers follow the master behavior by tracking desired relative positions to the leader while the leader attempts to travel along a given trajectory. It is assumed that most of the existing L-F approaches are implemented with a fixed topology. In many real-time applications such as source seeking and obstacle avoidance, shrinking or expanding the size of the formation pattern over time based on the spacing between obstacles is only the first step for guarantee of system safety. In case the distances between obstacles only fit one robot at a time, the robots should be able to switch to a column-line formation, in which robots move one after another. Although this problem was solved in \cite{Cao10,Hu12,Wen16, Dong16, Dong17}, the previous experimental results were used to only validate their theories while implementation of given methods in real-time applications, which also play an important role, have not been taken into account. Moreover, in \cite{Dong16} and \cite{Dong17}, time-varying formation control laws were implemented on the five UAVs with the pre-defined interaction topologies. In multiple obstacles avoidance schemes via queuing behavior, each edge representing communication links should be generated by robots so that the process of obstacles avoidance based on formation variation will take place smoothly and gently.

Most recent formation control methods have not given sufficient attention to obstacle and inter-robot collision avoidance in uncertain environments. In \cite{Dai14} and \cite{Dai15}, a collision-free way-point is determined based on the Geometric Obstacle Avoidance Control Method (GOACM). Although the real robot demonstrations were conducted successfully, this approach would not work with complex-shaped obstacles. In addition, when steering around obstacles and moving in close proximity to each other, it is essential to have active inter-vehicle collision avoidance. There are a number of suggested techniques. According to \cite{Hafner11}, the velocities of both vehicles within potential collision areas are adjusted accordingly. Although real-time experiments were conducted successfully when a possible collision between two vehicles is prevented, this method needs to be extended to collisions involving more than two vehicles, with a better brake and throttle velocity control algorithm. According to \cite{Abeywickrama17}, the Potential Field Method (PFM) is utilized for path planning and inter-UAV collision avoidance. Nevertheless, the repulsive force, which drives the UAV away from collisions, is produced by the gradient of the repulsive potential functions which is very hard to be determined in real applications.

Another issue is involved in the leader/follower-loss scheme. In L-F strategy, F is equipped with a sensor to detect the relative position from F to L. However, this device is generally subject to specific constraints, such as restricted Field Of View (FOV) or depth range. Besides, maintaining the observation between L and F can be challenging if the sensor's FOV is directly blocked by obstacles, or temporarily disabled by effects such as motion blur or sudden lighting change. In these situations, tracking of the leader can be lost and fault-tolerant strategies to regain visibility should be designed. There are some results aiming to maintain visibility between robots \cite{Léchevin06, Morbidi11, Panagou14}. However, some of the literatures do not consider the presence of obstacles while others assume that the environment information is perceived a priori. In \cite{Wang18}, the follower is driven to where the contact with the leader was lost. Nevertheless, the robots are incapable of avoiding mobile obstacles and inter-robot collision may occur if the leader stops at a signal-loss point or travels with a slower velocity. Being more reliable than the predictive technique, another approach using additional robotic vision sensors was developed in \cite{Sundvall06}. In this paper, two or more sources of location measurements are used to detect unexpected position deviation based on an extended Kalman filter. Unfortunately, the detector can malfunction if obstacles obstruct the robot camera's FOV. These constraints have been overcome in recent research using cooperation between air-ground vehicles, allowing the shortcomings of each type (payload, perception,  etc.) to be overcome by the complementary skills provided by the others \cite{Ariyur08, Michael14, Forster13}. For example, the UAV has a clearer view and is less obstructed than the UGVs and camera imagery from the UAV can assist the UGVs to plan a safer trajectory. A practical application for this cooperative hybrid system is to support ground transportation tasks with a low-altitude UAV flying above the UGVs \cite{Ariyur08, Guérin15}. In order to track the UGVs and provide camera imagery of sufficient quality for localization, precise control of the UAV is of foremost importance. However, the common control approaches, e.g. Proportional-Integral-Derivative (PID),  employed in small UAVs such as quadrotor and drones generally cannot meet the desired performance requirements while withstanding environmental disturbances \cite{Mellinger12}. Therefore, control of UAVs in the presence of large wind gusts, or aggressive maneuvers is required. Some recent studies have utilized on-board wind sensors to estimate the wind forces \cite{Perozzi17}. Nevertheless, these methods cannot measure the wind gusts' precise direction and add unwanted costs. Another contribution in this field is to model the wind velocity \cite{Schiano14,Sikkel16}, however, due to complete dependence on a precise model with no adaptation for nonparametric or uncertain parameters, it is still not an effective solution. Furthermore, an adaptive controller, which has the ability of conforming to any uncertainties in the dynamics caused by wind gusts, is illustrated in \cite{Fernández17}. Unfortunately, only an inner-loop velocity controller is designed and the effects of the wind disturbances are tested in only the vertical direction.

Compared to the previous results presented in the literature, the contributions of this study are threefold. Firstly, an NI-based obstacle and inter-collision avoidance control scheme, combined with an NI time-varying formation tracking control approach with switching communication topology is presented for distributed NI L-F non-economic multi-vehicles moving in hazardous environments. Thanks to this new technique, mobile robots can altogether change their functions within the L-F architecture and shift their positions in the formation to generate a new pattern. Secondly, collision avoidance between L and F is solved based on calculating the intersection area between virtual circles surrounding the robots to provide a virtual repelling force. Furthermore, a new NI distributed obstacle avoidance strategy is also integrated into each robot, allowing them to sense two collision scenarios (one obstacle, two arrays of multiple obstacles) and execute a proper methodology by themselves. Finally, a position tracking controller for the UAV is developed to alleviate the adverse effects of gusts. Providing a wider view from above, this UAV will attempt to keep the mobile UGVs within the field of view of its camera.

In order to conduct real-time experiments, it is assumed that measurement of relative positions between the leader and follower and between the robots and obstacles will be simulated using an indoor Motion Capture System (MCS). Furthermore, we simulate the relative position measuring capability of the UAV camera using our MCS to calculate relative position between the UAV and each UGV. Finally, unpredicted wind gusts are emulated by an electrical box fan which has 3 speed modes.

The remainder of this paper is organized as follows. Section II introduces the hardware/software configuration and architecture design for the collision  avoidance capability. Preliminaries on NI formation architecture are given in Section III. UAV and UGV dynamics are mentioned in Section IV. A trajectory tracking controller with disturbance rejection applied to UAV is analyzed and designed in Section V. The obstacle and inter-collision avoidance using NI theory is proposed in Section VI. The stability of the whole structure is proven in Section VII. The simulation results for the proposed approaches are illustrated in Section VIII, followed by the experiment results in Section IX. Conclusions are drawn in Section X.

\section{Experimental Apparatus}
\subsection{Hardware/Software Configuration}

The hardware setup used in this work consists of one UAV (AR Drone quadrotor platform), three heterogeneous UGVs (Pioneer P3-AT and P3-DX), one electrical fan and four obstacles (plastic packing boxes). Notably, the wheels of each UGV are equipped with optical encoders to estimate the linear velocity, moving distance and yaw angular rate. For software settings, two systems were employed for data collection and interaction protocol. All necessary data, including posture and relative position of agents and obstacles, are analyzed and measured by a Vicon Motion Capture System (MCS). Discrete commands or additional data exchanges among the robots as well as those between the ground station (GS) and the robots are achieved utilizing the Robot Operating System (ROS).  

\subsection{Overall Architecture}

\begin{figure}
	\begin{center}
		\includegraphics[width=0.75\linewidth]{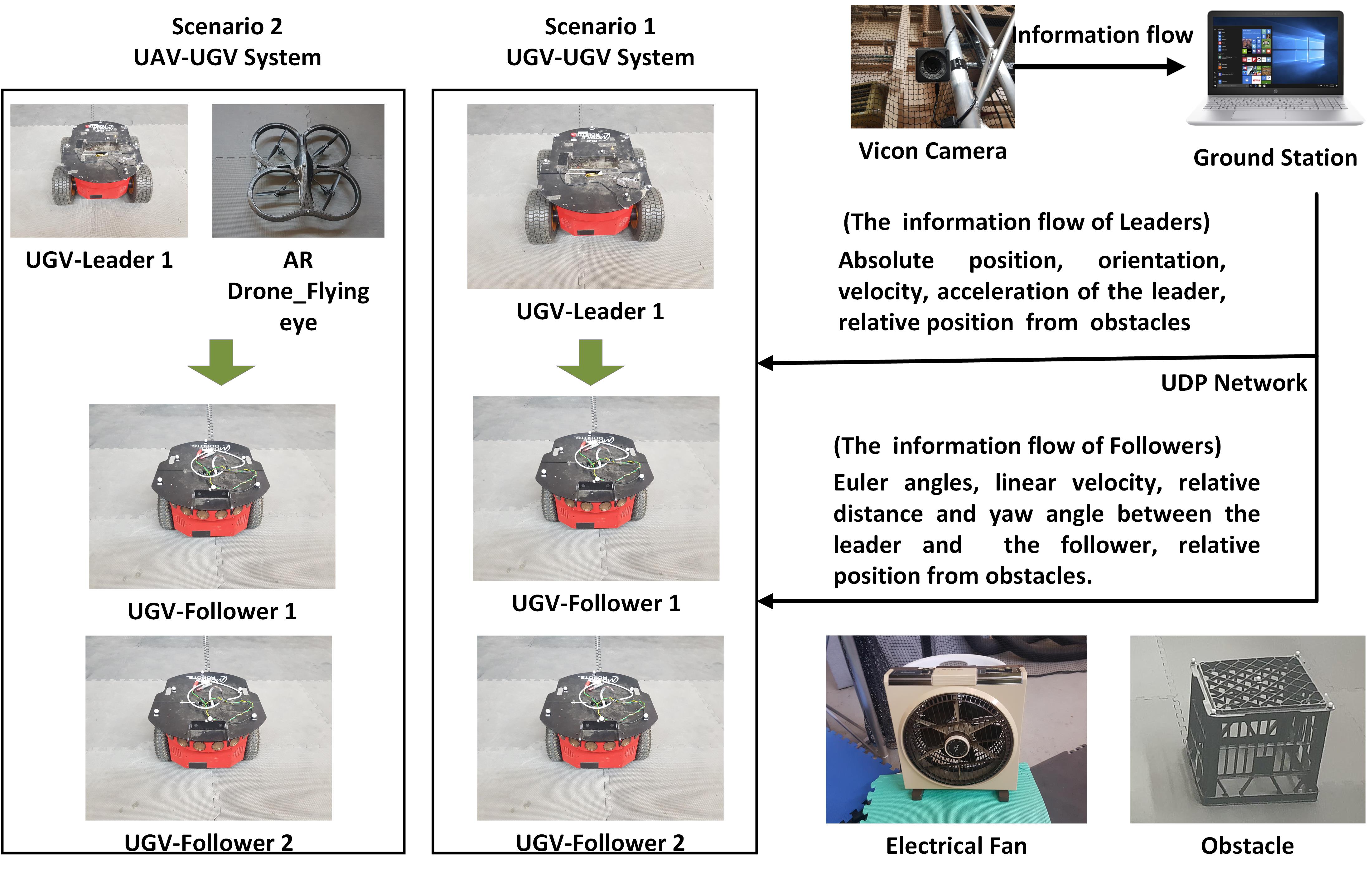}    
		\caption{Overall Architecture Diagram.}  
		\label{fig:architecture}                                 
	\end{center}                                 
\end{figure}

Our Leader-Follower (L-F) control network architecture is designed to perform cooperative scenarios with either UAV-UGV or UGV-UGV systems. The functionality and the task of each block are adequately described as shown in Fig. \ref{fig:architecture}. Data collected in the Vicon system is broadcasted continuously to each agent at a frequency of 100 Hz using a UDP network protocol.

\section{Preliminaries on Negative-Imaginary Formation Control Theory}

\begin{lemma}[Petersen, I.R. and Lanzon, A.,2010]\label{SNIsystem} 
In the single input/single output (SISO) case, a transfer function becomes strictly negative-imaginary (SNI) if all its poles have negative real parts and its Nyquist plot is contained below the real axis. According to \cite{Pertersen10}, this lemma can be mathematically defined as follows: \textit{P(s)} $\in$ \textit{SNI} if \textit{j[P(j$\omega$)-P$^{*}$(j$\omega$)] $>$ 0} for all \textit{$\omega$ $>$ 0}.
\end{lemma}

\begin{figure}
	\begin{center}
		\includegraphics[width=0.75\linewidth]{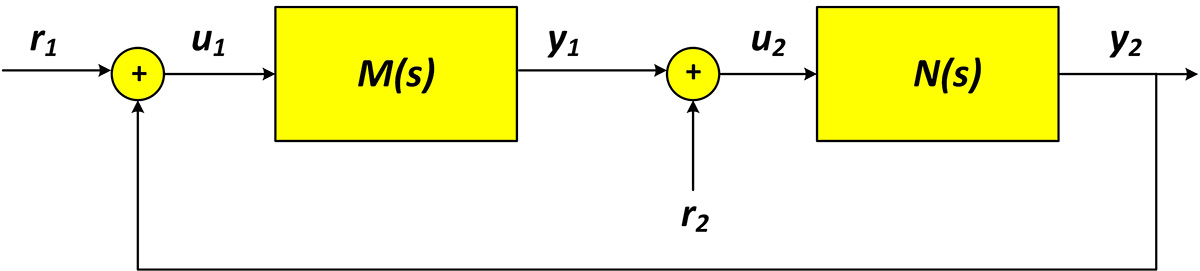}    
		\caption{A NI closed-loop control diagram.}  
		\label{fig:SNI_flow_chart}                                 
	\end{center}                                 
\end{figure}

Since an SNI plant is found, \cite{Pertersen10} and \cite{Xiong09} state that a necessary and sufficient condition for the internal stability of a positive feedback interconnection between an SNI plant with transfer function matrix \textit{M(s)} and an SNI/NI controller with transfer function matrix \textit{N(s)} (see Fig. \ref{fig:SNI_flow_chart}) is provided as follows:

\begin{equation}
\lambda_{max}(M(0)N(0)) < 1
\end{equation}

Let \textit{$\lambda_{max}(.)$} denote the maximum eigenvalue of transfer functions while \textit{r$_{i}$, u$_{i}$ and y$_{i}$} represents the reference, input and output signal.

\begin{figure}
	\begin{center}
		\includegraphics[width=0.8\linewidth]{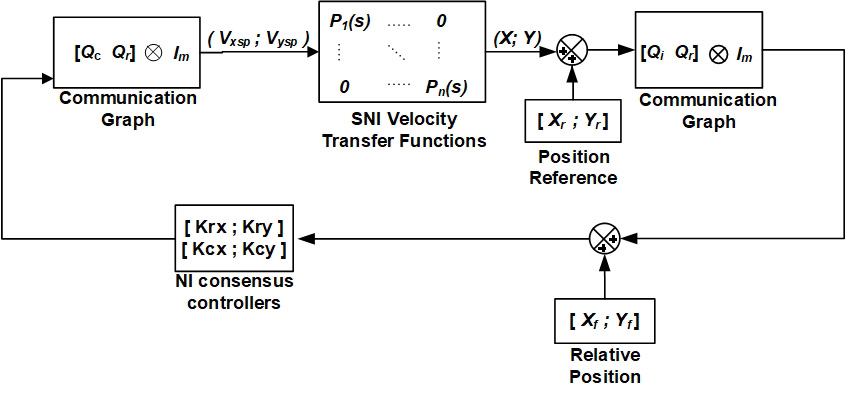}    
		\caption{A NI formation control protocol for multi-UAVs.}  
		\label{fig:UAV_consensus}                                 
	\end{center}                                 
\end{figure}

Developed from the consensus approach of MIMO NI systems \cite{Wang15},  an NI-systems formation implementation was successfully designed and tested for multi-UAV models in our previous work \cite{Tran17}. In that paper, new reference matrix \textit{Q$_{r}$}, consensus matrix \textit{Q$_{c}$} and the offset distance terms \textit{(X$_{f}$, Y$_{f}$)} between UAVs are added into the original consensus structure to solve the existing formation control problems as presented in Fig. \ref{fig:UAV_consensus}. The major experimental results verify that the formation pattern generated by UAVs is maintained during their rectangular movements. The overall equation of this architecture is given as:

\begin{equation}
X_{r} = 1_{n}\otimes{r},
\end{equation}
\begin{equation}
e_{i} = y_{i} + X_{r},
\end{equation}
\begin{equation}
\bar{y} = ([Q_{i}~Q_{r}]\otimes{I_{m}})e_{i},
\end{equation}
\begin{equation}
e_{f} = \bar{y} + X_{f},
\end{equation}
\begin{equation}
u= ([Q_{c}~Q_{r}]\otimes{I_{m}})diag_{i=1}^{n}{P_{s}(s)}e_{f}
\end{equation}

where \textit{n}, \textit{m} and \textit{l} represent the number of agents, the maximum output of each agents and the number of edges in the information graph. \textit{r} $\in\mathbb{R}^{2\times1}$ denotes the reference position on the plane of the master, \textit{X$_{r}$} $\in\mathbb{R}^{2n\times1}$ is the reference matrix, and \textit{x}$_{f}$ corresponds to the relative position between the slave UAVs and the master UAV in the configured formation. \textit{u} $\in\mathbb{R}^{m\times{1}}$ is the velocity set point input of each UAV on the x and y axes while the output \textit{y}  $\in\mathbb{R}^{m \times{1}}$ is the current position of each UAV. $\overline{u}\in\mathbb{R}^{lm\times1}$ and $\overline{y}\in\mathbb{R}^{lm\times1}$ are the input and output of overall network plant. \textit{e$_{r}$} is the error between the desired position of the master and its current position, while \textit{e$_{f}$} is the error between the desired relative position and the current relative position of each UAV.  $P_{s}(s)\in\mathbb{R}^{lm\times lm}$ is the group of SNI consensus and tracking controllers for the group of UAVs, and \textit{Q$_{i}$}, \textit{Q$_{c}$}, and \textit{Q$_{r}$} are the incidence, consensus and reference matrix of agents respectively.

\section{UAV/UGV Dynamic Models}
A PID controller is employed to help our systems not only achieve the desired vertical and horizontal velocities but also satisfy the necessary and sufficient conditions for an SNI system as presented in Lemma \ref{SNIsystem}. The AR Drone responds to attitude commands sent from the ground station over WiFi. Meanwhile, the UGV receives the yaw rate and speed commands from the GS. Using data from the Vicon MCS, we derived the Multi-Input Multi-Output (MIMO) transfer functions for the AR Drone and the UGV based on the ARMAX model which has the standard form as follows:
\begin{equation}
A(z)y(k)=B(z)u(k-n)+e(k)
\end{equation}

where \textit{(u(k),y(k))} are the system input and its output response, \textit{n} is the system delay, \textit{k} is the present time and \textit{e(k)} is the disturbance in the system.

\subsection{SNI Closed-Loop Transfer Function Models for a UAV}
The transfer function of the X-axis loop is given from the system identification as follows:
\begin{equation}\label{eqn:7}
 \frac{velxsp(s)}{posx(s)}= \frac{3.31s +  195.26}{s^2 + 174.66s + 3.12} 
\end{equation}
where \textit{velxsp(s)} and \textit{posx} highlight the actual position on the x axis and its desired velocity value. The vertical model in (\ref{eqn:7}) has damping factors of \textit{$\zeta_{1,2}$} = 1.  

The transfer function of the Y-axis loop can be elaborated as follows:

\begin{equation}\label{eqn:8}
\frac{velysp(s)}{posy(s)}= \frac{3.31s + 26.02}{s^2 + 25.71s + 0.18} 
\end{equation}
where \textit{velysp(s)} and \textit{posy} denote the actual position on the y axis and its desired velocity value. The horizontal model in (\ref{eqn:8}) gives \textit{$\zeta_{1,2}$} = 1.

\begin{figure}
	\begin{center}
		\includegraphics[width=0.8\linewidth]{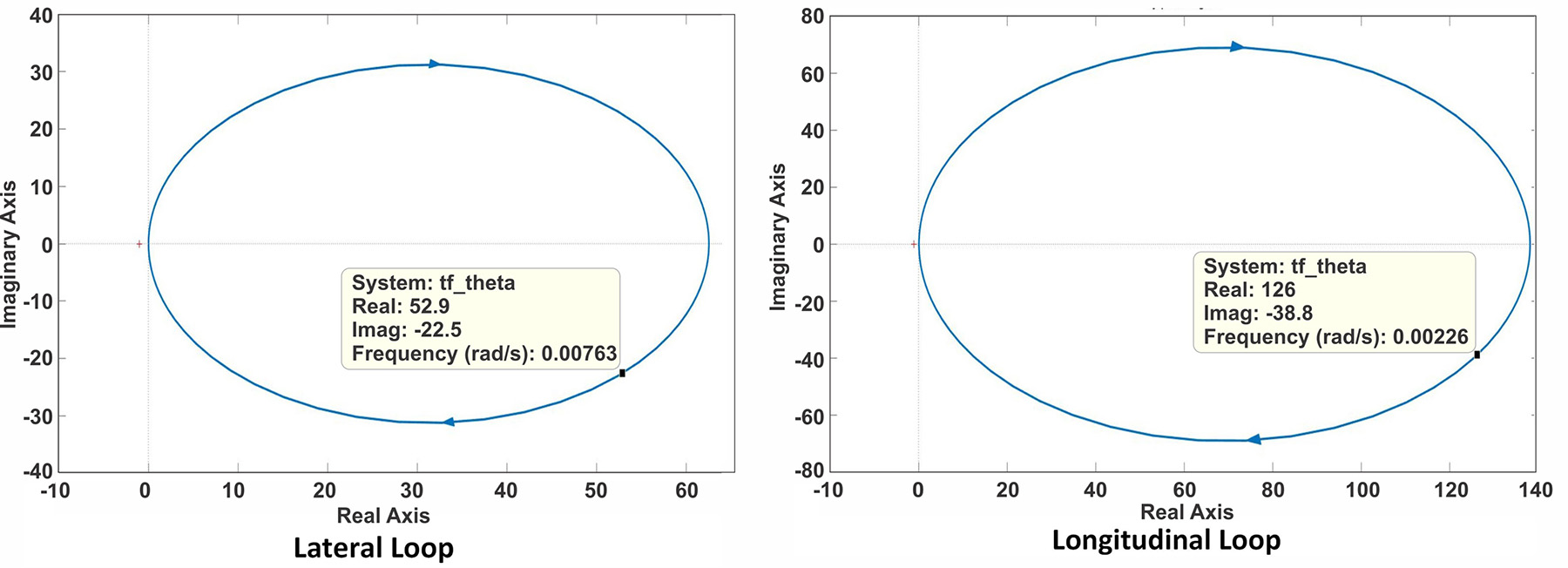}    
		\caption{The Nyquist plots of two velocity control loops for the UAV.}  
		\label{fig:UAV_SNI}                                 
	\end{center}                                 
\end{figure}

The Nyquist plots of two transfer functions are drawn in MATLAB. As seen in Fig. \ref{fig:UAV_SNI}, all found mathematical models satisfy the criteria of SNI systems.

\subsection{SNI Closed-Loop Transfer Function Models for a UGV}

\begin{figure}
	\begin{center}
		\includegraphics[width=0.8\linewidth]{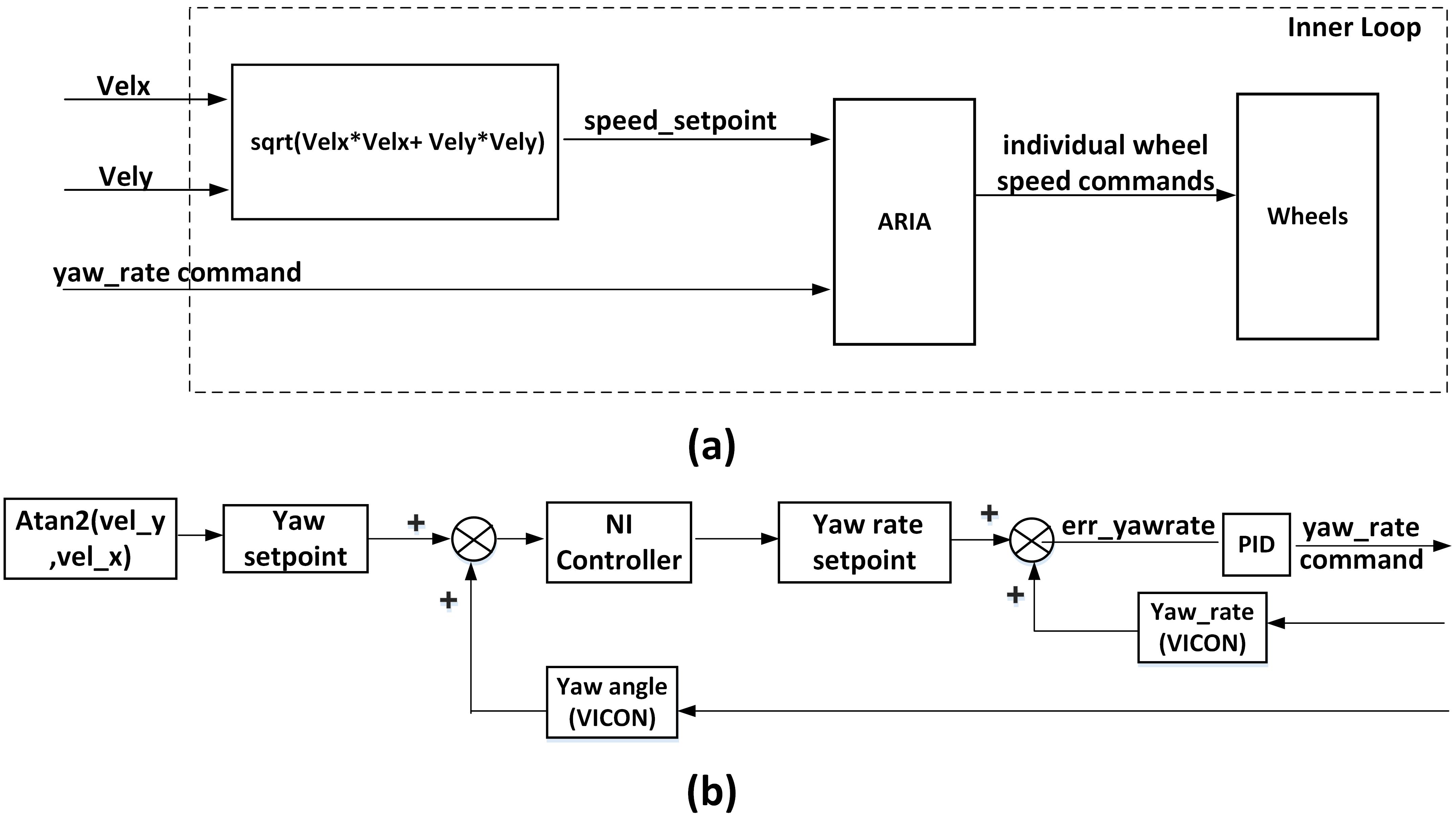}    
		\caption{The inner control diagram of a Pioneer P3-AT.}  
		\label{fig:x_y_velocity_UGV_update2}                                 
	\end{center}                                 
\end{figure}

The control input signals for the UGV consist of two inputs: yaw rate and speed. While the set-point translational velocities on the x and y axis drive the UGV to move to the target point, these need to be converted into yaw rate and speed commands (see Fig. \ref{fig:x_y_velocity_UGV_update2}). We first achieve this using a yaw control loop based on an NI-PID controller as shown in Fig. \ref{fig:x_y_velocity_UGV_update2}b. The desired yaw angle \textit{$\psi_{sp}$} is determined from the desired translational velocities as follows:
\begin{equation}
\psi_{sp} = atan2(vely,velx)
\end{equation} 

Based on the linear velocity on the x and y axis, the scalar speed command is achieved as follows:

\begin{equation}
V = \sqrt{vely^2+velx^2}
\end{equation}

From our systems identification, we obtain the following transfer function for the translational speed and yaw angle:
\begin{equation}\label{eqn:9}
\frac{dis(s)}{velsp(s)} = \frac{-0.15s^3+112.9s^2+4320.5s+1847912.3}{s^4 + 186.9s^3 + 58740s^2 + 1969445s + 39036.5}
\end{equation}
\begin{equation}\label{eqn:10}
\frac{\psi(s)}{\dot{\psi}_{sp}(s)} = \frac{17.25s^2-1018.48s+65838.57}{s^3 + 1401.1s^2 + 560049.64s + 68857.54}
\end{equation}
where \textit{$\psi$} and \textit{$\dot{\psi}_{sp}$} denotes the actual yaw output and its yaw angular rate reference signal. \textit{dis} presents the movement distance while a translational velocity \textit{velsp} is provided.

To provide better tracking on paths which have sudden turns at sharp corners, the position tracking controller is completely turned off and only the yaw controller is enabled to rotate the UGV around the z axis until the rotation is accomplished. 

\begin{figure}[h!]
	
	\centering
	\includegraphics[width=10pc, height=3cm]{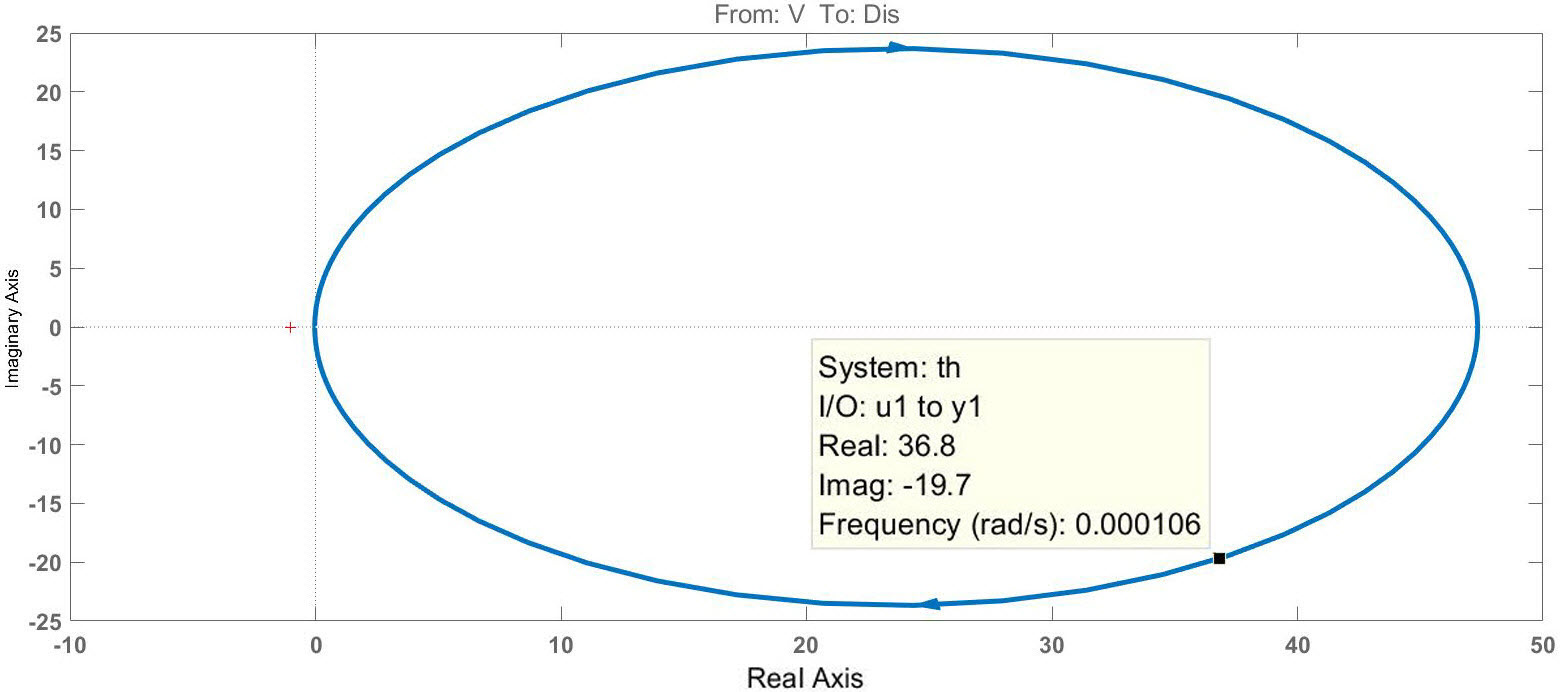}
	\includegraphics[width=10pc, height=3cm]{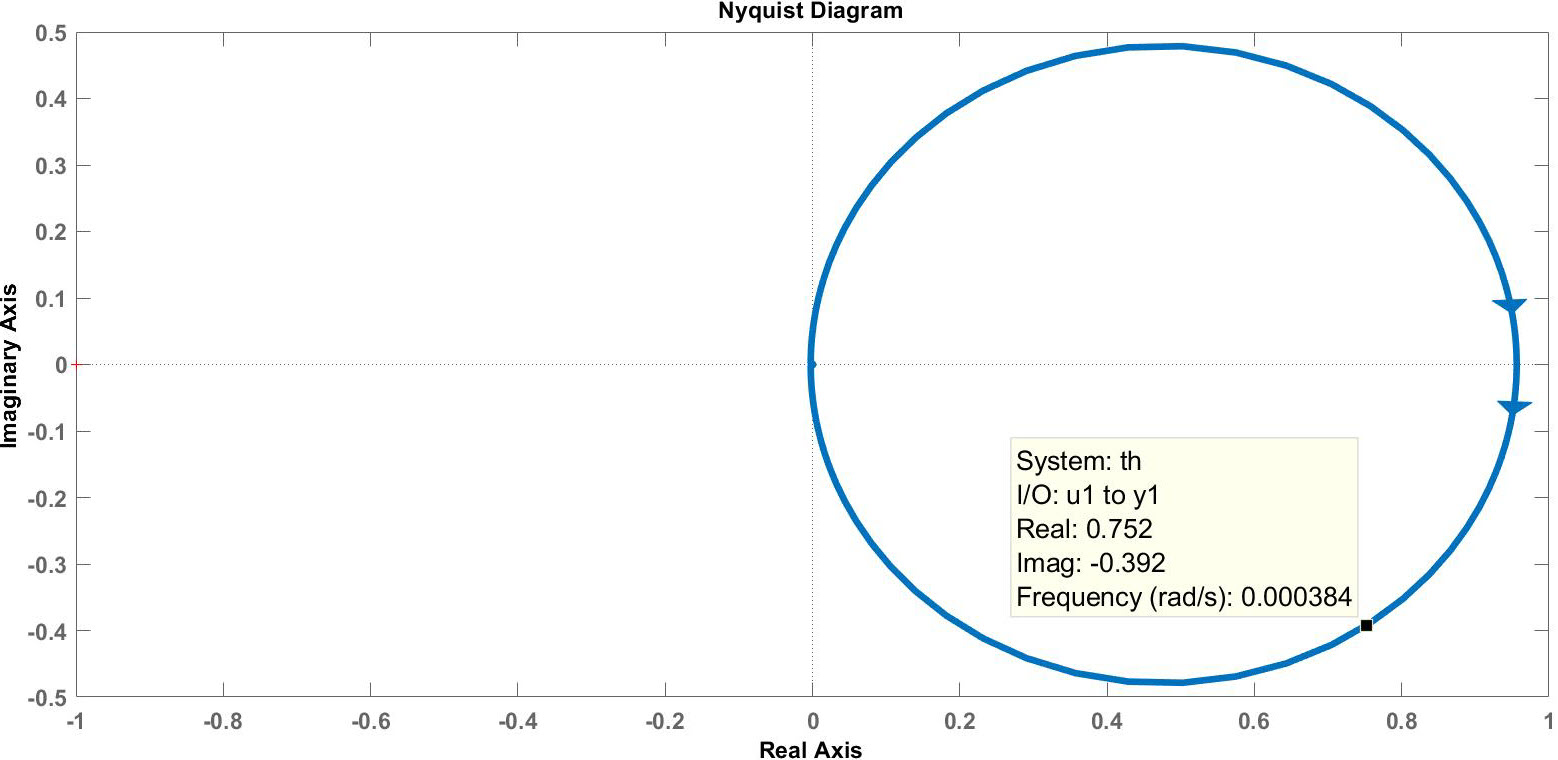}
	
	\caption{The Nyquist plot of the UGV velocity transfer function (left) - The Nyquist plot of the UGV yaw transfer function (right).}
	
	\label{fig:io}
	
\end{figure}

Based on Lemma \ref{SNIsystem}, the transfer functions found can be classified as SNI systems (see Fig. \ref{fig:io}).

\section{SNI trajectory tracking controller with external disturbance restriction for a UAV}
As discussed in Section I, the UAV is known to be sensitive to wind gusts.  Thus, it is vital to design a trajectory tracking controller which has good wind resistance characteristics. In this section, we propose two new approaches to solve this issue. While the first one is a new position control architecture, allowing the UAV system to obtain the expected position and velocity simultaneously while flying to its target, the second demonstrates the ability of our SNI controller to reject wind-gust disturbances.

\subsection{Position and Velocity Control Structure Using NI System Theory}

\begin{figure}
	\begin{center}
		\includegraphics[width=0.8\linewidth]{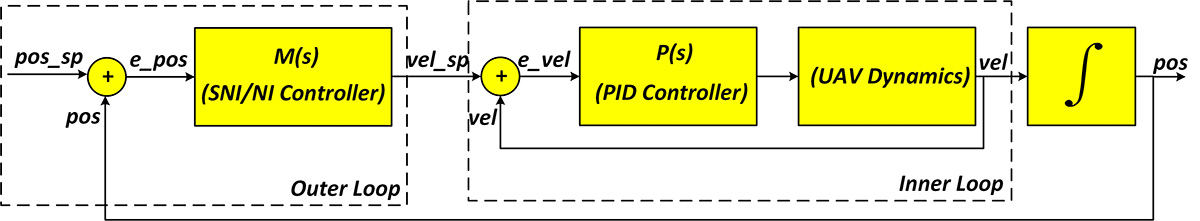}    
		\caption{Block diagram of the proposed control system for our
			quadrotor.}  
		\label{fig:UAV_Controller}                                 
	\end{center}                                 
\end{figure}

We separate the flight control problem into an inner loop that controls the velocity and an outer loop that controls the translational trajectory of the UAV. According to Fig. \ref{fig:SNI_flow_chart}, the NI feedback control structure has two input/output signals which are connected in series. Similarly, the overall block diagram of the control loops of our ARDrone is illustrated in Fig. \ref{fig:UAV_Controller}.

\begin{figure}
	\begin{center}
		\includegraphics[width=0.7\linewidth]{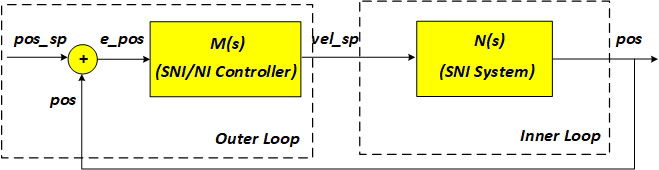}    
		\caption{Reduced block diagram of the proposed control system for our
			quadrotor.}  
		\label{fig:Reduced_UAV_Controller}                                 
	\end{center}                                 
\end{figure}

Noted that the inner velocity feedback control loops are already designed in Section IV, where PID controllers are implemented to track the velocity setpoints on the x-y plane. Besides, their closed-loop transfer functions also exhibit the SNI property. We outline a reduced control diagram as shown in Fig \ref{fig:Reduced_UAV_Controller}.

Under the control of an SNI controller, the output of outer loop provides velocity setpoint, which is regarded as the input signal for internal velocity loop while that of the inner produces a corresponding movement for UAV. The input/output relationships can be mathematically summarized as follows:
\begin{equation}
e\textunderscore pos = pos\textunderscore sp + pos
\end{equation}
\begin{equation}
vel\textunderscore sp = e\textunderscore pos*M(s)
\end{equation}
\begin{equation}
e\textunderscore vel = vel\textunderscore sp + vel
\end{equation}
\begin{equation}
vel = N(s)*e\textunderscore vel
\end{equation}
\begin{equation}
pos = \int_{0}^{t} vel dt
\end{equation}

where \textit{e\textunderscore pos} denotes the position derivation between the desired and actual position (\textit{pos\textunderscore sp} and \textit{pos}), while \textit{e\textunderscore vel} indicates the velocity error between the velocity output (\textit{vel}) and its reference signal (\textit{vel\textunderscore sp}).

The overshoot problem, that refers to the phenomenon in which the UAV has unexpected swing movement when reaching its target position due to movement inertia, is most likely to exist in recent studies on UAV that have been implemented with the position controller alone. Such problems can be practically eliminated in our control approach using the combination of the desired position and velocity controls, in which the UAV’s velocity is proportionally decreased when the distance between it and the target becomes shorter and shorter, and therefore it will take a shorter time for the UAV position to settle its position.

\subsection{SNI Controller for Outer Loop}
It is hard for a UAV system to achieve its desired performance due to high sensitivity to unmodeled or uncertain environmental effects. Therefore, active damping augmentation, in addition to the faster time response, used to counteract the effects of external commands and disturbances value, must be considered carefully. In order to satisfy these two conditions, we now employ the general form of the first-order SNI controllers from the desired position inputs (\textit{posx\textunderscore sp(t), posy\textunderscore sp(t)}) to the desired velocity outputs (\textit{velx\textunderscore sp(t), vely\textunderscore sp(t)}) given by
\begin{equation}\label{eq:16}
M(s) = \frac{\delta}{a s + \omega^2} = \frac{\delta/\omega^2}{a/\omega^2 s + 1} = \frac{K}{\tau s + 1} 
\end{equation}
where, \textit{$a$} $>$ 0 is the viscous damping constant relative to \textit{$\omega$} $>$ 0. \textit{$\delta_{i}$} points out the gain values of our controller. The transfer function in (\ref{eq:16}) gives the time constant \textit{$\tau$} = $\frac{a}{\omega^2}$ and the DC gain \textit{K} = $\frac{\delta}{\omega^2}$.
\begin{proof}
Let \textit{$\omega$} $>$ 0, \textit{$\delta$} $>$ 0 and \textit{$\tau$} $>$ 0, note that 
\begin{equation}
M(s) - M^T(−s) = \frac{\delta}{a s + \omega^2} - \frac{\delta}{-a s + \omega^2} = \frac{2\delta s}{a^2s^2 - \omega^4}
\end{equation}
has no zeros on the imaginary axis except at possibly s = 0.  It then follows from Lemma \ref{SNIsystem} that the outer-loop model in (\ref{eq:16}) is an SNI controller.
\end{proof}

It should be pointed out that the higher the assigned value \textit{$\delta$} is, the faster the system response is. If the gain increases to a sufficiently high threshold, the system can become unstable. \textit{$\tau$} and \textit{$K$} must be adapted accordingly to drive the system back to equilibrium. 

\subsection{Motion Planning for a UAV}

\begin{figure}
	\begin{center}
		\includegraphics[width=0.65\linewidth]{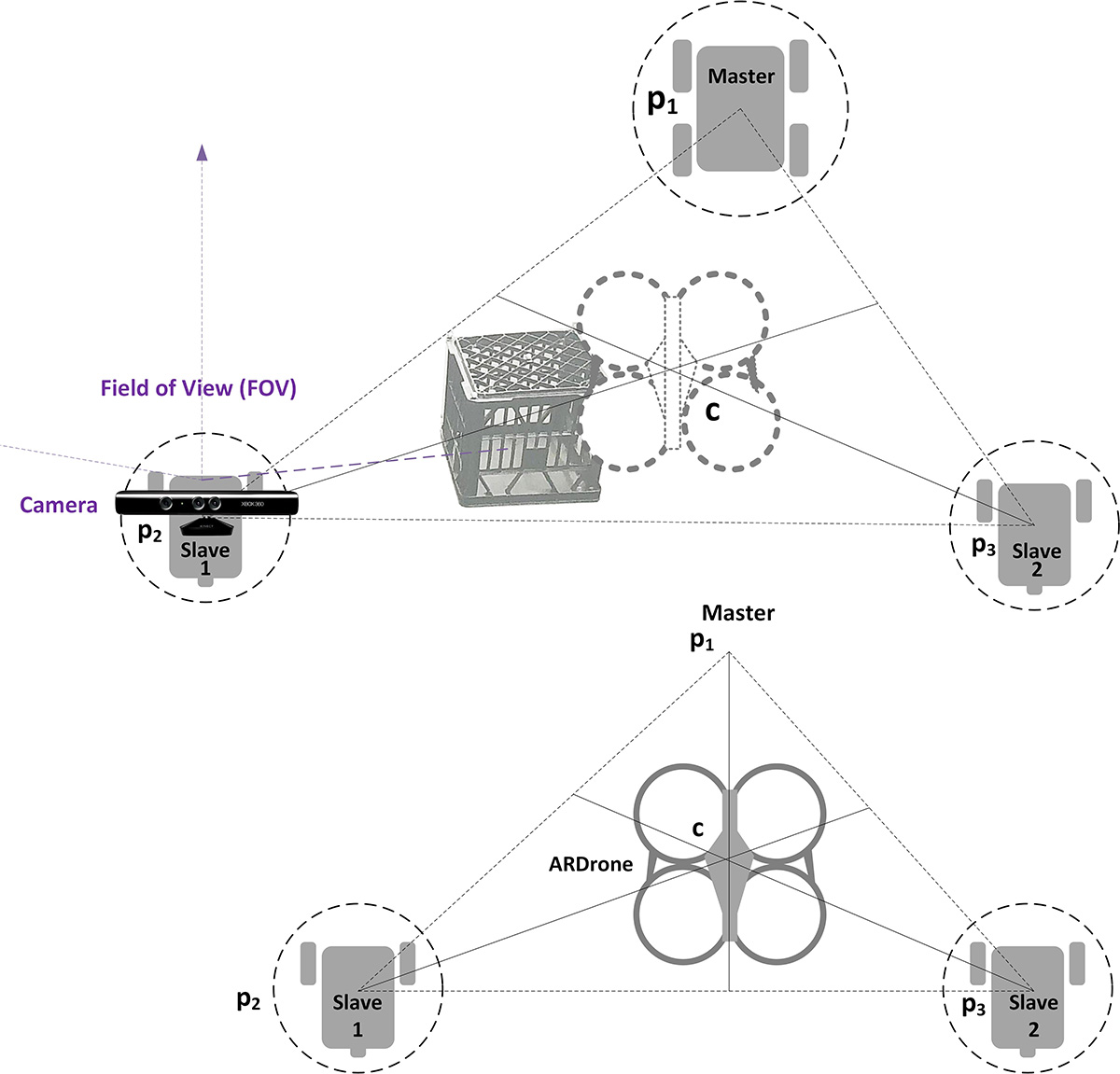}    
		\caption{An air-ground cooperative approach for the leader/follower UGVs loss situation. In the picture, the relative distances with the remaining robots cannot be measured by Slave 1's camera because its FOV is completely obscured by one obstacle. However, they can be easily processed by the UAV's camera.}  
		\label{fig:UAV-center}                                 
	\end{center}                                 
\end{figure}

A motion planning system generally generates a discrete-waypoint path from a starting location to a destination taking into account the robot's motion constraints. As mentioned earlier, a potential solution for when an obstacle obscures a UGV camera is studied in this paper. The UAV is controlled by a new position controller to follow the geometrical center-point of the formation pattern as presented in Fig \ref{fig:UAV-center}. We formulate the center-point coordinate as follows:
\begin{equation}
c_{i} = \frac{\sum_{j=1}^{n} pos_{ij}}{n} \forall i \in \{1..2\}
\end{equation}
where \textit{i} illustrates the horizontal or vertical axis, \textit{n} denotes the number of UGVs on the ground, and \textit{pos$_{ij}$} presents the UGVs location on the x and y axes.

Thanks to this technique, a redundant measurement of distances between UGVs is obtained by the camera installed on the underside of the UAV. Once the leader or follower state measured by its camera is lost, a command inside the UGV is triggered to utilize the data computed from the UAV camera instead.

\section{Distributed Obstacle and Inter-Collision Avoidance Control via a Distributed L-F Approach}

In this section, we would like to handle the first and second issues by introducing a distributed L-F control scheme, a distributed obstacle avoidance method, and a mutual collision avoidance approach, which means that algorithms are run in each agent instead of a ground station. 

\begin{figure}
	\begin{center}
		\includegraphics[width=0.36\linewidth]{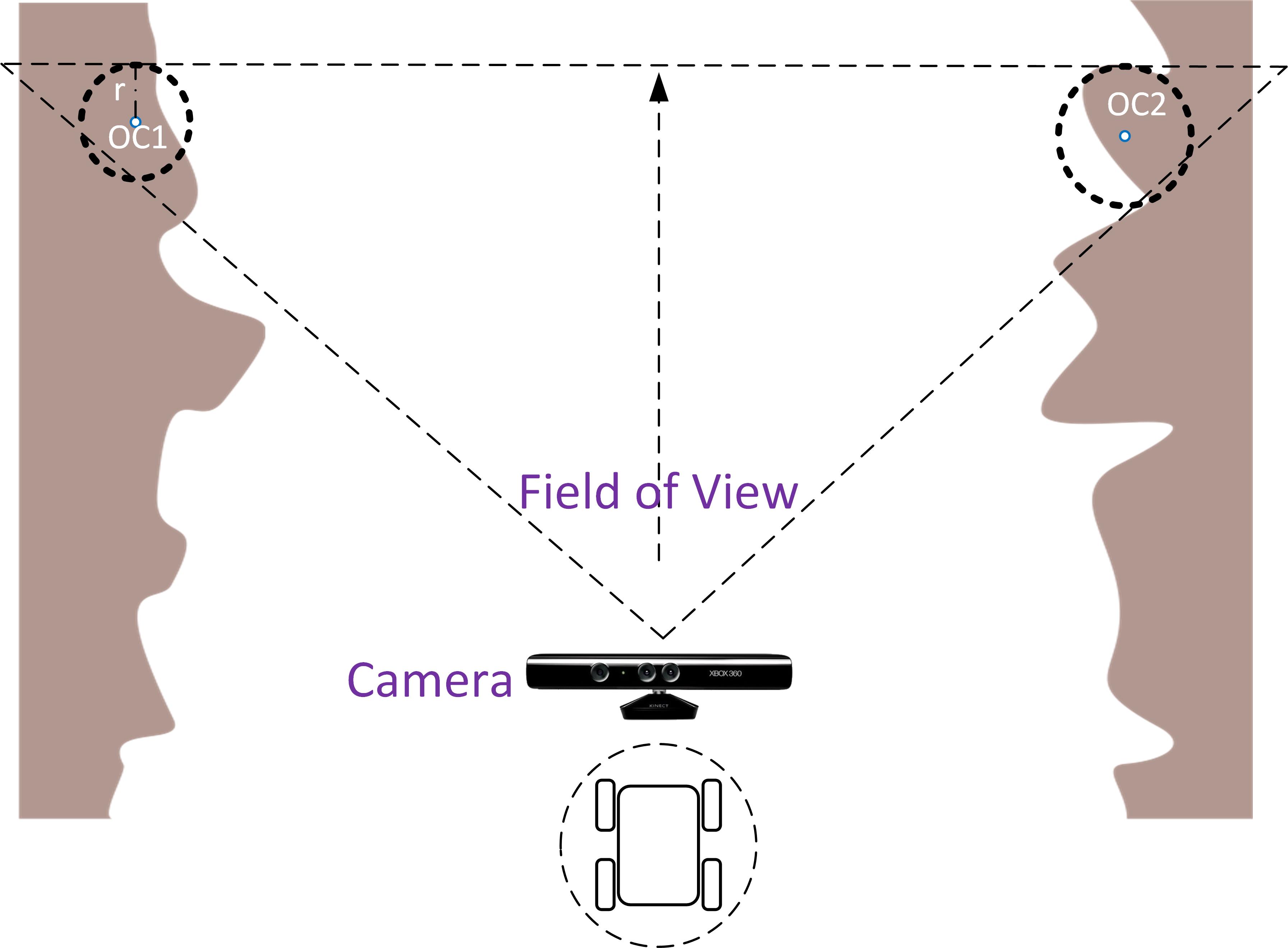}    
		\caption{Obstacle detection algorithm within FOV cameras.}  
		\label{fig:identification_obstacle}                                 
	\end{center}                                 
\end{figure}

Besides, it is assumed that the real shape of an obstacle is often irregular in an uncertain environment. Therefore, it is challenging to compute the escape angle by using the obstacles' actual boundary as per the traditional GOACM method does \cite{Dai14,Dai15}. To solve this issue, we generate a virtual circle that surrounds the recognized obstruction range within the camera FOV (see Fig. \ref{fig:identification_obstacle}). Its centroid is formulated by the geometric decomposition method (dividing the whole detected obstacle into a finite number of simpler figures). Its radius is determined by the distance from the center point to the camera's max view distance (from (\ref{eq:19}) to (\ref{eq:21})).

\begin{equation}\label{eq:19}
\epsilon_{i{x}}= \frac{\sum_{i}^{k}\epsilon_{i{x}}}{k}, \epsilon_{i{y}}= \frac{\sum_{i}^{k}\epsilon_{i{y}}}{k}
\end{equation}
\begin{equation}\label{eq:20}
\epsilon_{x}= \frac{\sum_{i}^{n}\epsilon_{i{x}}A_{i}}{\sum_{i}^{n}A_{i}}, \epsilon_{y}= \frac{\sum_{i}^{n}\epsilon_{i{y}}A_{i}}{\sum_{i}^{n}A_{i}}
\end{equation}
\begin{equation}\label{eq:21}
r_{x} = \epsilon_{x} + \max (FOV_{x}),r_{y} = \epsilon_{y} + \max (FOV_{y})
\end{equation}

where \textit{k} is a specific number of simpler patterns within a recognized obstacle area. \textit{$\epsilon_{i}$} and \textit{A$_{i}$} indicate the centroid location and area of each part. \textit{r$_{i}$} is the obstacle virtual circle's radius.

\subsection{Distributed L-F Strategy of Mobile Robots}
The distributed L-F strategy can be regarded as a process of automatically defining the robots' role via the relative distance between the robots location and its destination as well as between L and F, and then moving to the corresponding position in a known formation. These computations take place inside the UAV's built-in computer. When this task is accomplished, identification (ID) numbers that stipulate the title of each UGVs in their formation are published to corresponding followers within UAV's camera FOV.

\begin{figure}
	\begin{center}
		\includegraphics[width=0.72\linewidth]{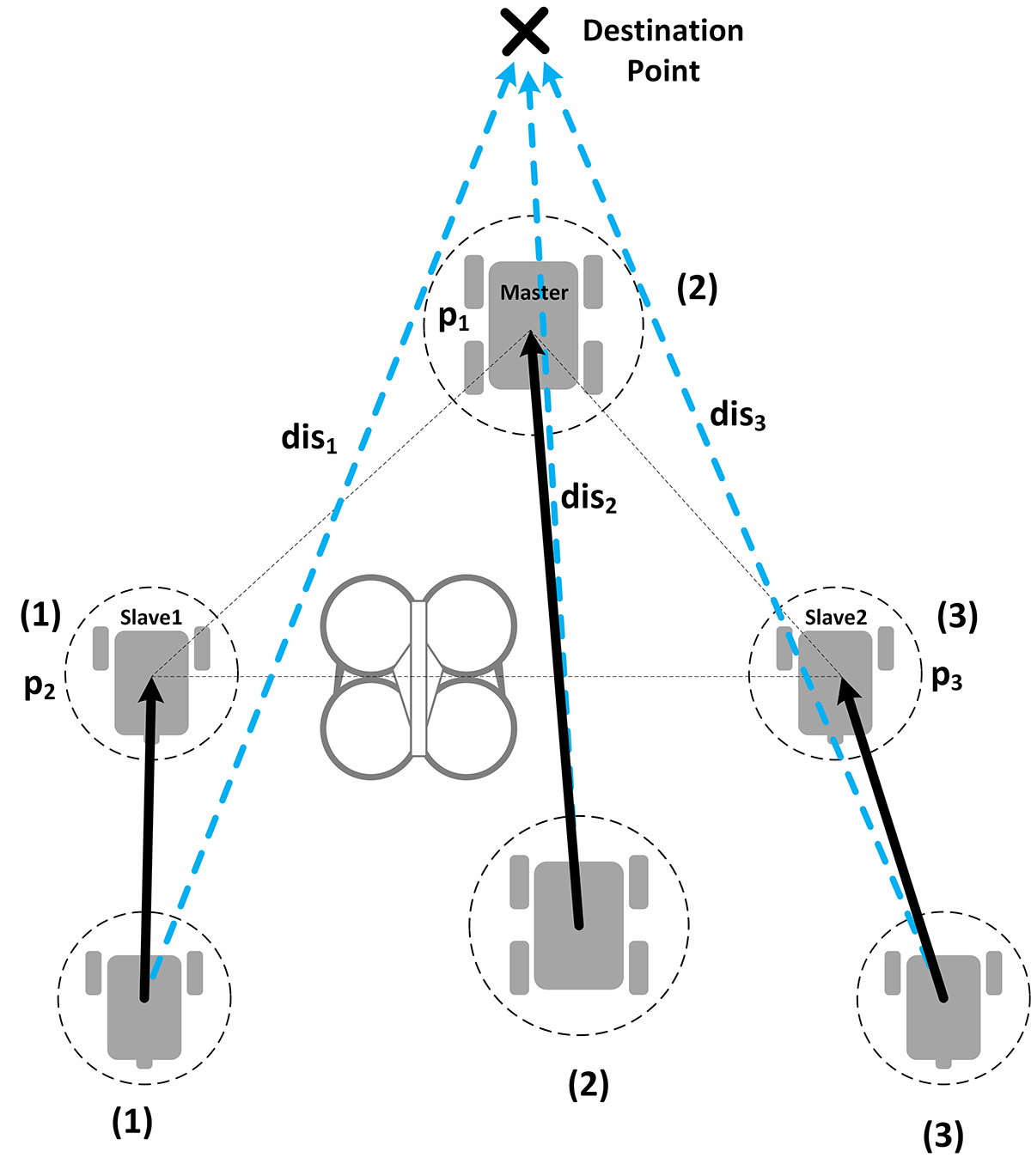}    
		\caption{Desired formation is generated by UGVs using distributed L-F strategy.}  
		\label{fig:Distributed_L_F}                                 
	\end{center}                                 
\end{figure}

For example, Fig \ref{fig:Distributed_L_F} shows that UGV (2) among the three UGVs has the shortest distance from the destination point; therefore, it is defined as L and assigned ID as (1). Next, (1) is closer to L's left-side position; thus, it becomes F1 and is labeled ID 2. This process is repeated until the final robot is numbered. We elaborate rules of distributed L-F strategy in a matrix \textit{$ID_{1\times n}$} as follows:
\begin{equation}
ID =
\begin{cases}
ID_{i} = 1 &\text{if } i|dis_{di} < dis_{dj} \forall i,j \in \{1..n\} \& i\neq j, \\
ID_{i} = 0 &\text{Otherwise}.
\end{cases}
\end{equation}
\begin{equation}
ID =
\begin{cases}
ID_{i} = k &\text{if } i|dis_{Li} < dis_{Lj} \forall i,j \in \{1..n\} \& i\neq j \\
		 & \& i\neq L \& k = k+1, \\
0 &\text{Otherwise}.
\end{cases}
\end{equation}
where \textit{dis$_{di}$} denotes relative distance between robots position and the destination while \textit{dis$_{Li}$} illustrates relative distance between L and F.

Regarding followers, a formation matrix \textit{$R_{1\times n}$} is configured in each UGV as below:
\begin{equation}
R = [0 \quad dis_{L{1}} \quad dis_{L{2}} \quad ... \quad dis_{L{n}}]
\end{equation}

where \textit{dis$_{L{1}}$}, \textit{dis$_{L{2}}$}, ... , \textit{dis$_{L{n}}$} present the desired relative distances between L and F in their formation. As a result, the ID number received will correspond with the column index of the \textit{R} matrix.

\subsection{Distributed Obstacle Avoidance Control Scheme with Dynamic Interactive Topology Switches}

Although in our previous work \cite{Tran17}, the NI formation architecture was presented for only controlling an invariant formation, that architecture can also be applied to modify the robot's formation pattern over time. Therefore, we develop this theory to solve the multiple-obstacles avoiding problem based on the NI consensus-based formation control protocol.

We recall the concept of the NI formation control algorithm, employed in each robot. Considering the input-output data space \textit{z} = [vel\_sp$^{(t)}$  pos$^{(t)}$],  the recursive procedure for our method can be described as follows:
\begin{equation}
vel\_sp =\begin{bmatrix}
V_{xsp_{L}}^{(t)}\\
V_{ysp_{L}}^{(t)}\\
V_{xsp_{F_{1}}}^{(t)}\\
V_{ysp_{F_{1}}}^{(t)}\\
\vdots\\
V_{xsp_{F_{n-1}}}^{(t)}\\
V_{ysp_{F_{n-1}}}^{(t)}\\
\end{bmatrix}
=\begin{bmatrix}
~\text{Kr$_{x}$$\times$(X$_{r}$ + X$_{L}$)} \\
~\text{Kr$_{y}$$\times$(Y$_{r}$ + Y$_{L}$)} \\
~\text{Kc$_{x_{1}}$$\times$(disx$_{r}$+disx$_{LF_{1}}$)}\\
~\text{Kc$_{y_{1}}$$\times$(disy$_{r}$+disy$_{LF_{1}}$)}\\
\vdots\\
~\text{Kc$_{x_{n-1}}$$\times$(disx$_{r}$+disx$_{LF_{n-1}}$)}\\
~\text{Kc$_{y_{n-1}}$$\times$(disy$_{r}$+disy$_{LF_{n-1}}$)}\\
\end{bmatrix}
\end{equation}
\begin{equation}
pos =\begin{bmatrix}
pos_{x_{L}}^{(t)}\\
pos_{y_{L}}^{(t)}\\
pos_{x_{F_{1}}}^{(t)}\\
pos_{y_{F_{1}}}^{(t)}\\
\vdots\\
pos_{x_{F_{n-1}}}^{(t)}\\
pos_{y_{F_{n-1}}}^{(t)}\\
\end{bmatrix}
= \overline M(s)\times vel\_sp
\end{equation}
where \textit{(X$_{r}$, Y$_{r}), ($X$_{L}$,Y$_{L}$)} are the desired and actual position on the x and y axis while \textit{(disx$_{r}$, disy$_{r}$), (disx$_{LF}$, disy$_{LF}$)} are the desired and actual distance between L and F. \textit{(Kr$_{x}$,Kr$_{y}$), (Kc$_{x}$, Kc$_{y}$)} are the SNI/NI consensus controllers of L and F as illustrated in Section V.

A time-varying formation can be achieved in a shorter time interval if \textit{(Kr$_{x}$,Kr$_{y}$, (Kc$_{x}$,Kc$_{y}$))} are assigned with higher values.
\begin{equation}
k=\begin{bmatrix}
Kr_{x}\\
Kr_{y}\\
Kc_{x_{i}}\\
Kc_{y_{i}}\\
\vdots\\
Kc_{x_{n-1}}\\
Kc_{y_{n-1}}\\
\end{bmatrix}
=\begin{bmatrix}
~\text{disNO$_{x_{L}}$/[t$\times$(X$_{r}$ + X$_{L}$)]} \\
~\text{disNO$_{y_{L}}$/[t$\times$(Y$_{r}$ + Y$_{L}$)]} \\
~\text{disNO$_{x_{F_{i}}}$/[t$\times$(disx$_{r}$+disx$_{LF_{i}}$)]}\\
~\text{disNO$_{y_{F_{i}}}$/[t$\times$(disy$_{r}$+disy$_{LF_{i}}$)]}\\
\vdots\\
~\text{disNO$_{x_{F_{n-1}}}$/[t$\times$(disx$_{r}$+disx$_{LF_{n-1}}$)]}\\
~\text{disNO$_{y_{F_{n-1}}}$/[t$\times$(disy$_{r}$+disy$_{LF_{n-1}}$)]}\\

\end{bmatrix}
\end{equation}

where \textit{(dis$_{NO_{x}}$,dis$_{NO_{y}}$)} are the desired distance between the expected and current L/F position in varying formation. \textit{t} is the desired time interval to transform from a former formation pattern to the new one.

When the spacing between two obstacles is less than that of two robots, the only way for the robots to pass through the obstacles is to change their formation shape. Thus, we outline a newly distributed formation variation control algorithm, which guides all robots to autonomously re-arrange into a line formation before passing through tightly spaced obstacles.

\begin{figure}
	\begin{center}
		\includegraphics[width=0.75\linewidth]{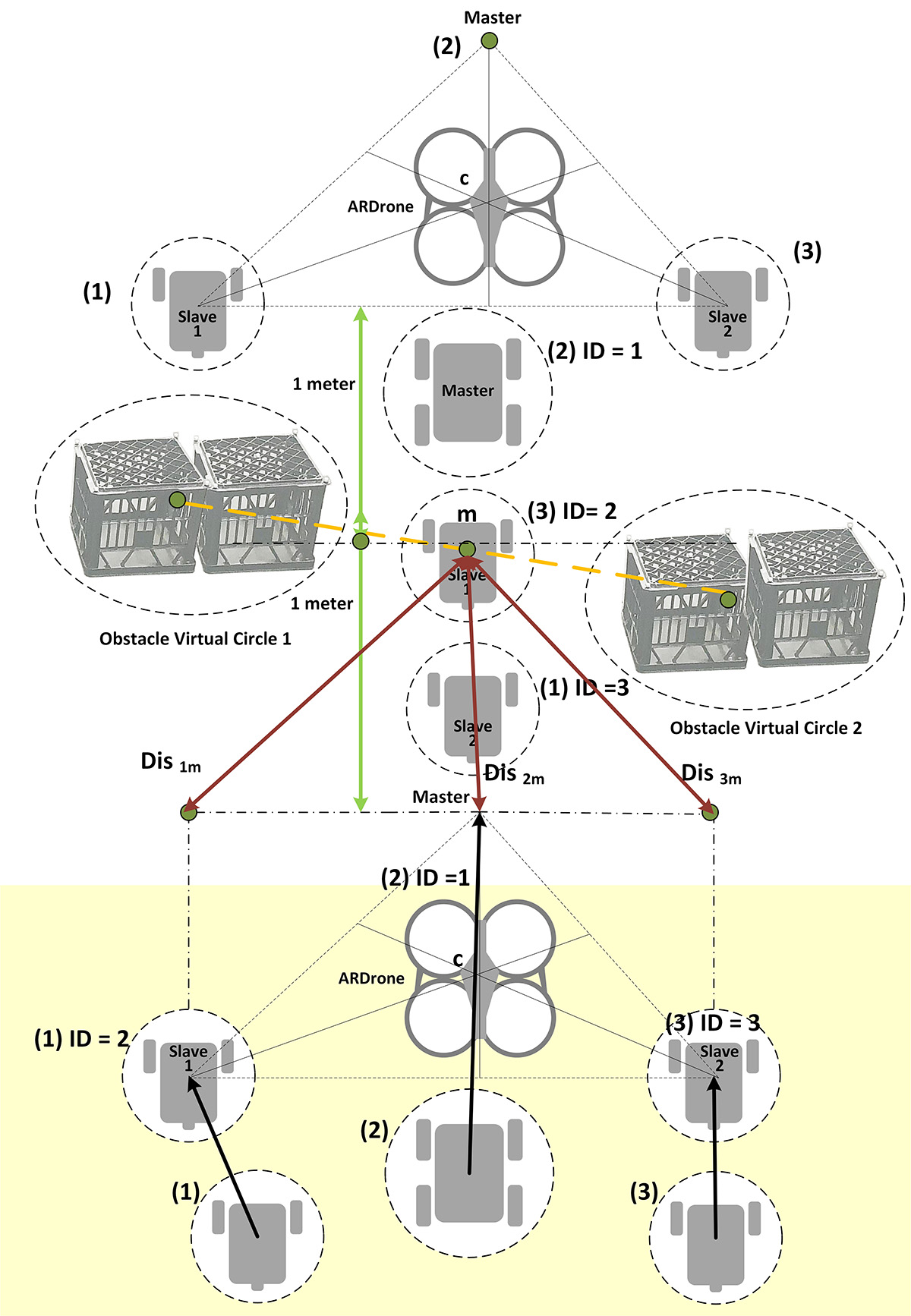}    
		\caption{Formation variation for three UGVs from a triangular pattern to a line-column one to avoid multiple-obstacles using NI obstacle avoidance strategy.}  
		\label{fig:queing_formation}                                 
	\end{center}                                 
\end{figure}

In this technique, the virtual midpoint of the connecting line \textit{m}, which is linking the obstacle's center points, is adopted to redetermine the role of each mobile robots in their formation by seeking out the smallest distance between \textit{m} and robots, and guide each robot to reach its proper position. This mode is activated while the distance from L or F to \textit{m} measured in each ground vehicle is less than 1 meter in front of \textit{m}, and is disabled while the distance from L or F to \textit{m} is larger than 1 meter behind \textit{m} as illustrated in Fig \ref{fig:queing_formation}.

\begin{equation}\label{eq:28}
que =\begin{cases}
que_{i} = 1 &\text{if } Dis_{im} < 1 \forall i \in \{1..n\}, \\ 
que_{i} = 0 &\text{if } Dis_{im} > 1 \forall i \in \{1..n\}.
\end{cases}
\end{equation}

When the value of \textit{que} in each robot becomes 1, ID numbers are recomputed based on the distance from \textit{m} to its location.

\begin{equation}\label{eq:29}
ID =
\begin{cases}
ID_{i} = k &\text{if } Dis_{im} < Dis_{jm} \forall i,j \in \{1..n\} \& i\neq j \\
		   & \& k = k+1, \\
ID_{i} = 0 &\text{Otherwise}.
\end{cases}
\end{equation}

This updated information is also known as the \emph{order reference number} among UGV followers. The desired position of the rear F UGV is always referred to the robot position in front of it with an offset distance while that of L on the horizontal axis is \textit{m}.

For example, Fig. \ref{fig:queing_formation} shows that there are two arrays of multiple obstacles blocking the moving path of three UGVs. Once (\ref{eq:28}) is satisfied, the UGV order adopted in the ID matrix is recomputed by comparing the distance from each robot to point \textit{m} (\ref{eq:29}). In this example, they are changed from (2)-(1)-(3) into (2)-(3)-(1). As a result, all UGVs will immediately generate a new line-column formation within the desired time interval based on the NI variant formation control algorithm and the dynamic ID interaction topology. After passing obstacles' midpoint \textit{m} for a distance of approximately one meter, each UGV in this collaborative group will go back to its former position in their prototype formation with the original interaction topology (2)-(1)-(3).

\subsection{Distributed Mutual-Collision Avoidance Control Scheme}

For the UGV to safely move around, the capabilities of localization, obstacle avoidance, inter-vehicle collision avoidance and the movement to target are
essentially needed. Such abilities ensure that the UGVs navigate a safe path and avoid collisions with obstacles while trying to reach their
goal. These fundamental functions are independently operated and not intimately connected to one another. Considering this aspect, we introduce a distributed mutual collision avoidance method between UGVs.

\begin{figure}
	\begin{center}
		\includegraphics[width=0.8\linewidth]{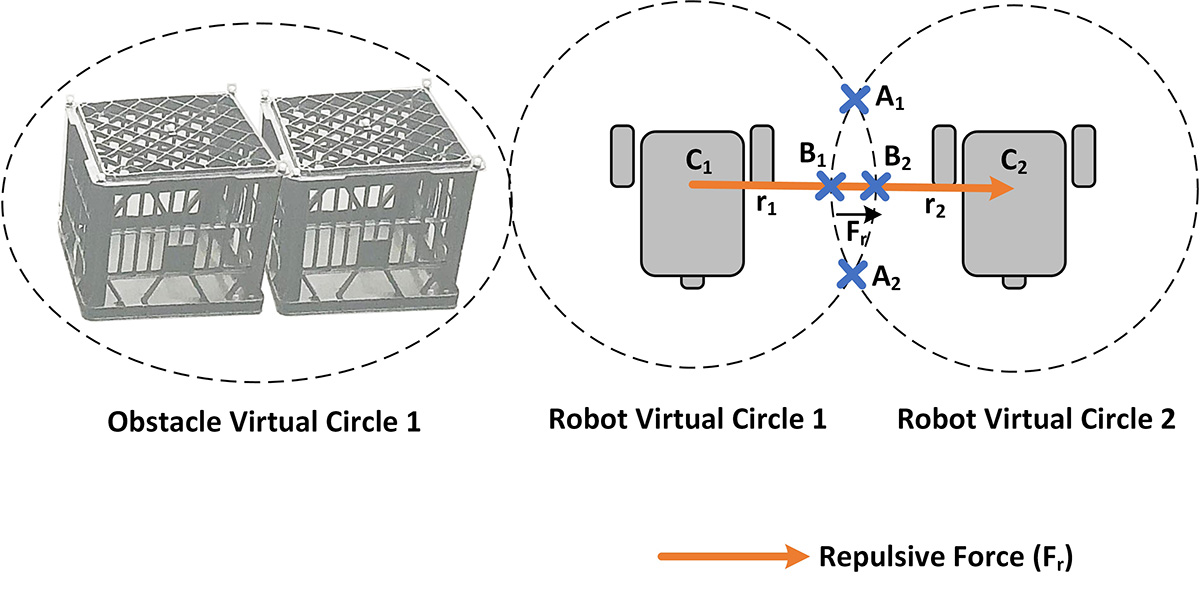}    
		\caption{Distributed NI inter-vehicle collision avoidance control method.}  
		\label{fig:Repulsive_force}                                 
	\end{center}                                 
\end{figure}

In our approach, each UGV is surrounded by a safe virtual circle. It is assumed that a mutual collision may occur only in case one of them are executing the obstacle avoidance behavior. While the safety margin of two robots is violated, a repulsive force \textit{F$_{r}$} is created to push the robot which has a free-collision path away. This force's magnitude and direction are determined as shown in Fig. \ref{fig:Repulsive_force}.

\begin{equation}
B_{1}B_{2} = C_{2}B_{1} - C_{2}B_{2},
\end{equation}
\begin{equation}
B_{1}B_{2} = r_{2} - (C_{1}C_{2} - C_{1}B_{2}),
\end{equation}
\begin{equation}
B_{1}B_{2} = r_{2} - (C_{1}C_{2} - C_{1}B_{2}),
\end{equation}
\begin{equation}
B_{1}B_{2} = r_{2} - (C_{1}C_{2} - r_{1}),
\end{equation}
\begin{equation}
F_{r} = k_{r}\times B_{1}B_{2}
\end{equation}

According to the Newtons' second law of motion, the acceleration set point for inter-vehicle collision avoidance is obtained as follows
\begin{equation}
\vec{a_{rsp}} = \frac{\vec{F_{r}}}{m}
\end{equation}
\begin{equation}
\vec{a_{rsp}} = \frac{k_{r}\times \vec{B_{1}B_{2}}}{m}
\end{equation}
\begin{equation}\label{eq:37}
P(s) = \frac{\vec{v_{rsp}}}{\vec{B_{1}B_{2}}} =  \frac{k_{r}}{ms}
\end{equation}
where \textit{m} is the mass of UGV and \textit{$\vec{a_{rsp}}$,$\vec{v_{rsp}}$} are the desired
avoiding acceleration and velocity on the x-y planar plane. \textit{r$_{1}$} and \textit{r$_{2}$} are radii of robot virtual circles. \textit{k$_{r}$} represents the stiffness of a virtual spring used to generate the repulsive force.

\begin{lemma}[Wang, J., Lanzon, A., Petersen, I. R.,2015]\label{NIsystem} 
The free body dynamics whose poles are at the origin and P($\infty$) = 0 are NI plants.
\end{lemma}

The main results of Lemma \ref{NIsystem} then implies that the transfer function of the mutual collision avoidance velocity given in (\ref{eq:37}) is an NI plant containing a single integrator (having a pole at the origin).

In order to simultaneously perform three fundamental tasks and create an NI uniform architecture, time-varying formation velocity and mutual-collision avoidance velocity are combined together with a priority weight for each task. For instance, in case the slave 1 has to avoid the remaining robot, the final equation is given as follows:
\begin{equation}
\begin{bmatrix}
V_{xsp_{L}}^{(t)}\\
V_{ysp_{L}}^{(t)}\\
V_{xsp_{F_{1}}}^{(t)}\\
V_{ysp_{F_{1}}}^{(t)}\\
\vdots\\
V_{xsp_{F_{n-1}}}^{(t)}\\
V_{ysp_{F_{n-1}}}^{(t)}\\
\end{bmatrix}
=\begin{bmatrix}
~\text{Kr$_{x}$$\times$(X$_{r}$ + X$_{L}$)} \\
~\text{Kr$_{y}$$\times$(Y$_{r}$ + Y$_{L}$)} \\
~\text{Kc$_{x_{1}}$$\times$[a$_{x1}$$\times$ (disx$_{r}$+disx$_{LF_{1}}$) + a$_{x2}$$\times$v$_{rxsp}$]}\\
~\text{Kc$_{y_{1}}$$\times$[a$_{y1}$$\times$ (disy$_{r}$+disy$_{LF_{1}}$) + a$_{y2}$$\times$v$_{rysp}$]}\\
\vdots\\
~\text{Kc$_{x_{n-1}}$$\times$(disx$_{r}$+disx$_{LF_{n-1}}$)}\\
~\text{Kc$_{y_{n-1}}$$\times$(disy$_{r}$+disy$_{LF_{n-1}}$)}\\
\end{bmatrix}
\end{equation}

where \textit{a$_{x1}$, a$_{x2}$, a$_{y1}$, a$_{y2}$} are the task priority weights respectively.

\section{Stability Proof of The Distributed Time-Varying Formation Control and Inter-Vehicles Collision Avoidance}

In this section, we recall the three concepts for SNI/NI MIMO systems presented in \cite{Wang15} and \cite{Pertersen10}.
\begin{lemma}[Wang, J., Lanzon, A., Petersen, I. R.,2015]\label{MIMOsystem} 
$\overline N(s)$ is SNI if and only if each member N(s) is SNI.
\end{lemma}
\begin{lemma}[Wang, J., Lanzon, A., Petersen, I. R.,2015]\label{MIMOStability} 
Given any SNI MIMO system $\overline M(s)$ and NI/SNI MIMO controller $\overline N(s)$, we obtain a stability result for the whole structure as follows
\begin{equation}
\lambda_{max}(\overline M(0)) \lambda_{max}(\overline N(0)) < \frac{1}{ \lambda_{max}(\mathcal{L})} = \frac{1}{ \lambda_{max}(QQ^{T})}
\end{equation}
\begin{equation}
\lambda_{max}(\overline M(0)) < \frac{1}{ \lambda_{max}(QQ^{T})\times \lambda_{max}(\overline N(0)))}
\end{equation}
\end{lemma}
\begin{lemma}[Petersen, I.R. and Lanzon, A.,2010]\label{SNINIsystem} 
	A positive connection between an NI system and an SNI system results in an SNI systems structure.
\end{lemma}

As proven in Section IV, each velocity transfer function N(s) represented UAV and UGV systems is SNI; hence, their $\overline M(s)$  plant is SNI according to Lemma \ref{MIMOsystem}. Consequently, our NI time-varying formation control architecture is stable if and only if the inequality in Lemma \ref{MIMOStability} is satisfied. 

On the other hand, \textit{P(s)} is NI, $\overline M(s)$ is SNI, a positive connection between \textit{P(s)} and  $\overline M(s)$ brings to an SNI structure $\overline E(s)$ as presented in Lemma \ref{SNINIsystem}.  

Since $\mathcal{L}$ and $\overline E(0)$ have fixed values which are greater than 0, only their SNI/NI controllers $\overline M(0)$ can be tuned to achieve the stability for time-varying formation control. For this reason, we select the negative gain values for $\overline M(s)$ so that the required stability is always guaranteed.

For example, regarding the UGV's velocity transfer function, their real DC gains at the zero frequency are equal to $\frac{1847912.3}{39036.5}$ = 47.34 [1]. In addition, $\mathcal{L}$ $>$ 0 [2].  All SNI controllers are chosen as  $\frac{-1}{s+1}$; therefore, its DC gain at the zero frequency is -1 [3]. Based on [1]-[3], the stability criteria in Lemma \ref{MIMOStability} is naturally satisfied.

\section{Computer Simulations}
Two cases are given to illustrate the main results of our theories. The first case examines the proposed approaches in Section VI on 6 UGVs, while the second case is to present the position tracking capability with the new structure mentioned in Section V. Its effectiveness is then compared to that of the traditional PID controller in the same structure.

\subsection{Distributed NI Collision Avoidance Approaches}
For the purpose of computer simulations, we have developed a Simulink model in the m-file environment, containing the real UGV dynamic models and our solutions to handle the first and second issue. 

For our computations, we use the following numerical data: \textit{m} = 1 kg, \textit{K$_{r}$} = \textit{K$_{c}$} = \textit{k$_{r}$} = -0.1, \textit{Fmax$_{r}$} = 6 N, \textit{Vmax$_{sp}$} = 2 cm/s, \textit{C$_{1}$} = [-65, 50] cm, \textit{C$_{2}$} = [50, -50] cm, \textit{r$_{1}$} = \textit{r$_{2}$} = 46 cm, \textit{r$_{ob}$} = 35 cm, target = [300, 450] cm. The initial position and velocity of UGVs are defined by two functions: 160$\times$(2$\times$rand([n,2])-1) and 0.002$\times$(rand([n,2])-350) respectively.

\begin{figure}
	\begin{center}
		\includegraphics[width=0.9\linewidth]{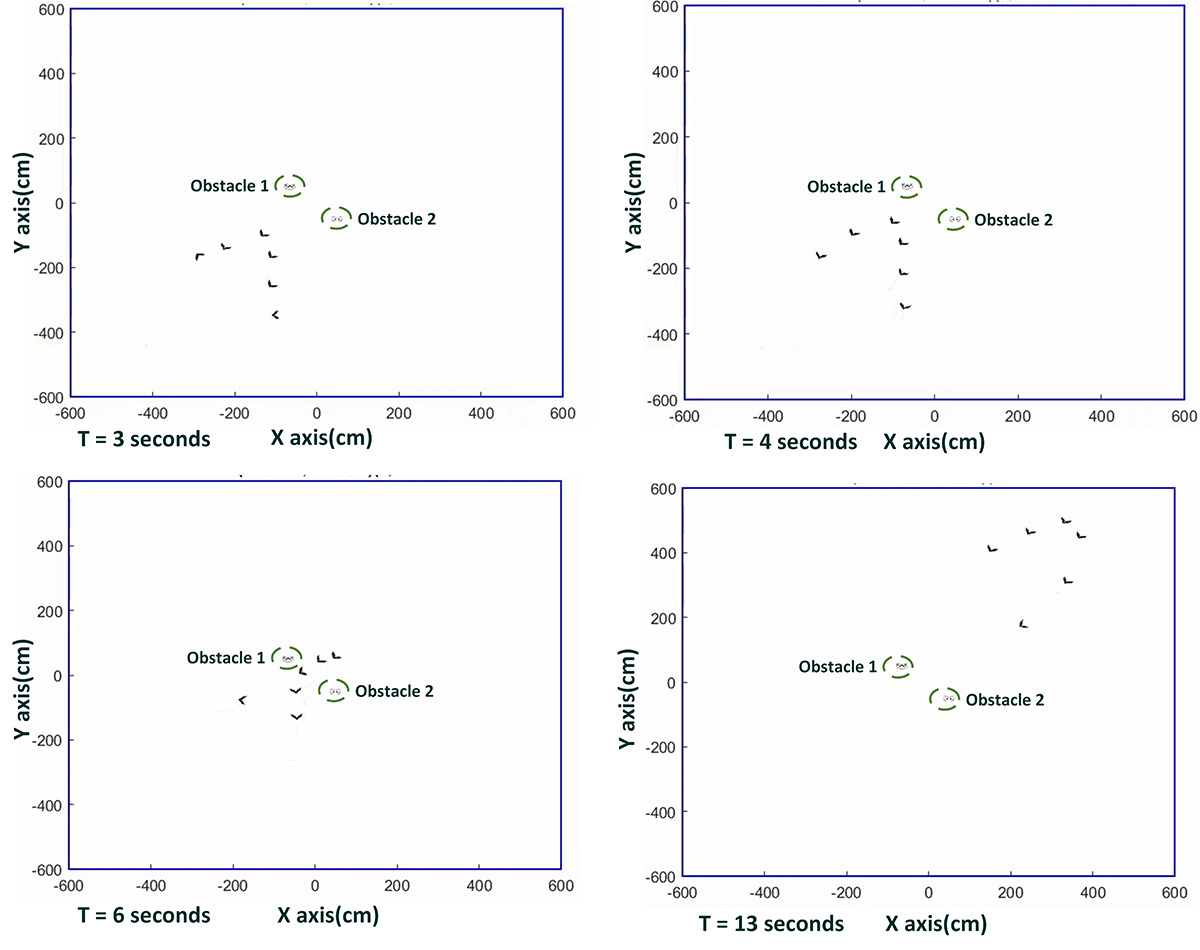}    
		\caption{Distributed NI time-varying formation and inter-vehicle avoidance control with dynamic topologies switches and distributed L-F strategy for 6 UGVs to generate the formation and avoid two arrays of multiple obstacles via the queuing behavior.}  
		\label{fig:result_sim1}                                 
	\end{center}                                 
\end{figure} 

\begin{figure}
	\begin{center}
		\includegraphics[width=0.87\linewidth]{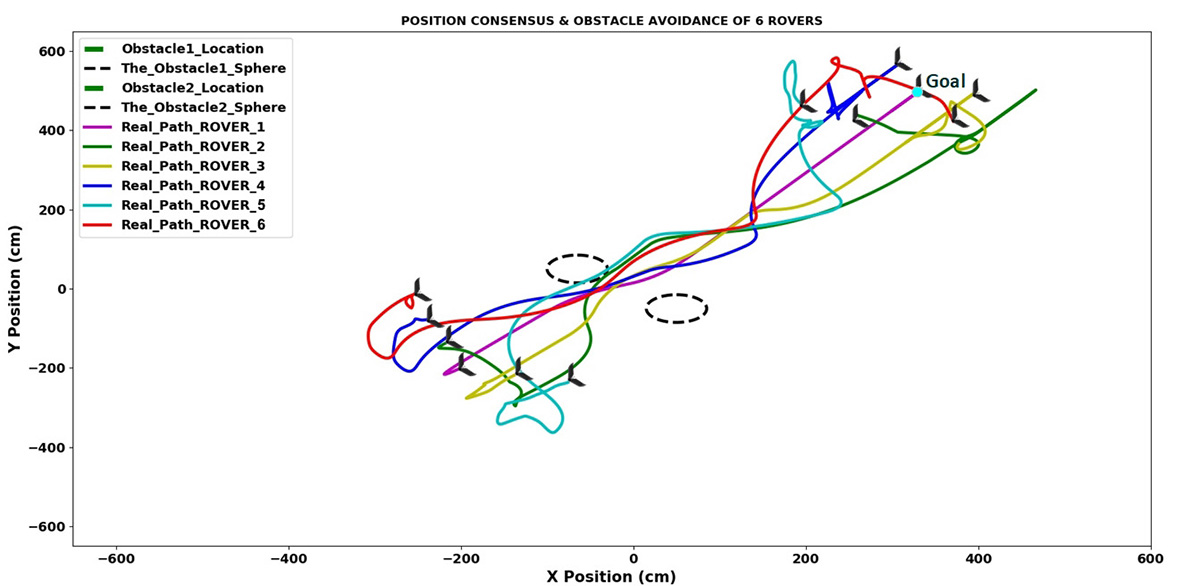}    
		\caption{Obstacles avoiding and target reaching paths involving 6 UGVs.}  
		\label{fig:data4}                                 
	\end{center}                                 
\end{figure} 

The performances of 6 UGVs are shown in Fig. \ref{fig:result_sim1} and Fig. \ref{fig:data4}, where a unit step position reference signal at target location is applied. As can be seen, with distributed NI collision avoidance approaches, the mobile robots can self-arrange their position in a pre-defined V-shaped formation and reach the target safely although their initial positions are random, and their traveling paths are hindered by two facing obstacles.

\subsection{Two Loops Position Tracking Controller for a UAV}
We first conduct extensive computer simulations to reflect the performance of the closed-loop control system. Based on the velocity transfer functions found in Section IV, we explore the ability of the SNI-PID controller while stabilizing the dynamics of our quadrotor platform in its two loops, namely, vertical, and horizontal. 

A unit step position reference signal with \textit{x$_{sp}$} = 0.5 m and  \textit{y$_{sp}$} = 0.5 m are produced at \textit{t} = 1s. The general form of SN-PID controller is expressed as below:

\begin{equation}
M\textunderscore x_{SNI}(s)= M\textunderscore y_{SNI}(s)= \frac{velx\textunderscore sp}{e\textunderscore posx}= \frac{vely\textunderscore sp}{e\textunderscore posy}= \frac{-1}{s+1}
\end{equation}
\begin{equation}
\begin{split}
P\textunderscore x(s)&= P\textunderscore y(s)= \frac{roll\textunderscore sp}{e\textunderscore xvel}= \frac{pitch\textunderscore sp}{e\textunderscore yvel}\\
&= -\frac{Kp_{vel}s+Ki_{vel}+Kd_{vel}s^{2}}{s}\\
&= -\frac{0.3162s+0.0021+0.135s^{2}}{s}
\end{split}
\end{equation}

where \textit{(roll\textunderscore sp, pitch\textunderscore sp)} are the Euler angle reference signals for the UAV. \textit{(Kp$_{vel}$, Ki$_{vel}$, Kd$_{vel}$)} are PID parameters for velocity control loop.

In order to validate the performance of SNI controller, we replace the SNI controller in the outer loop with the traditional PIDF controllers and repeat this procedure. The controller transfer functions are obtained as follows:

\begin{equation}
M\textunderscore x_{PIDF}(s)=\frac{velx\textunderscore sp}{e\textunderscore posx}= -\frac{0.0031s+0.000064+0.028s^{2}}{s^{2}+0.055s}
\end{equation}
\begin{equation}
M\textunderscore y_{PIDF}(s)=\frac{vely\textunderscore sp}{e\textunderscore posy}= -\frac{0.0611s+0.002+0.26s^{2}}{s^{2}+0.469s}
\end{equation}

\begin{figure}
	\begin{center}
		\includegraphics[width=0.85\linewidth]{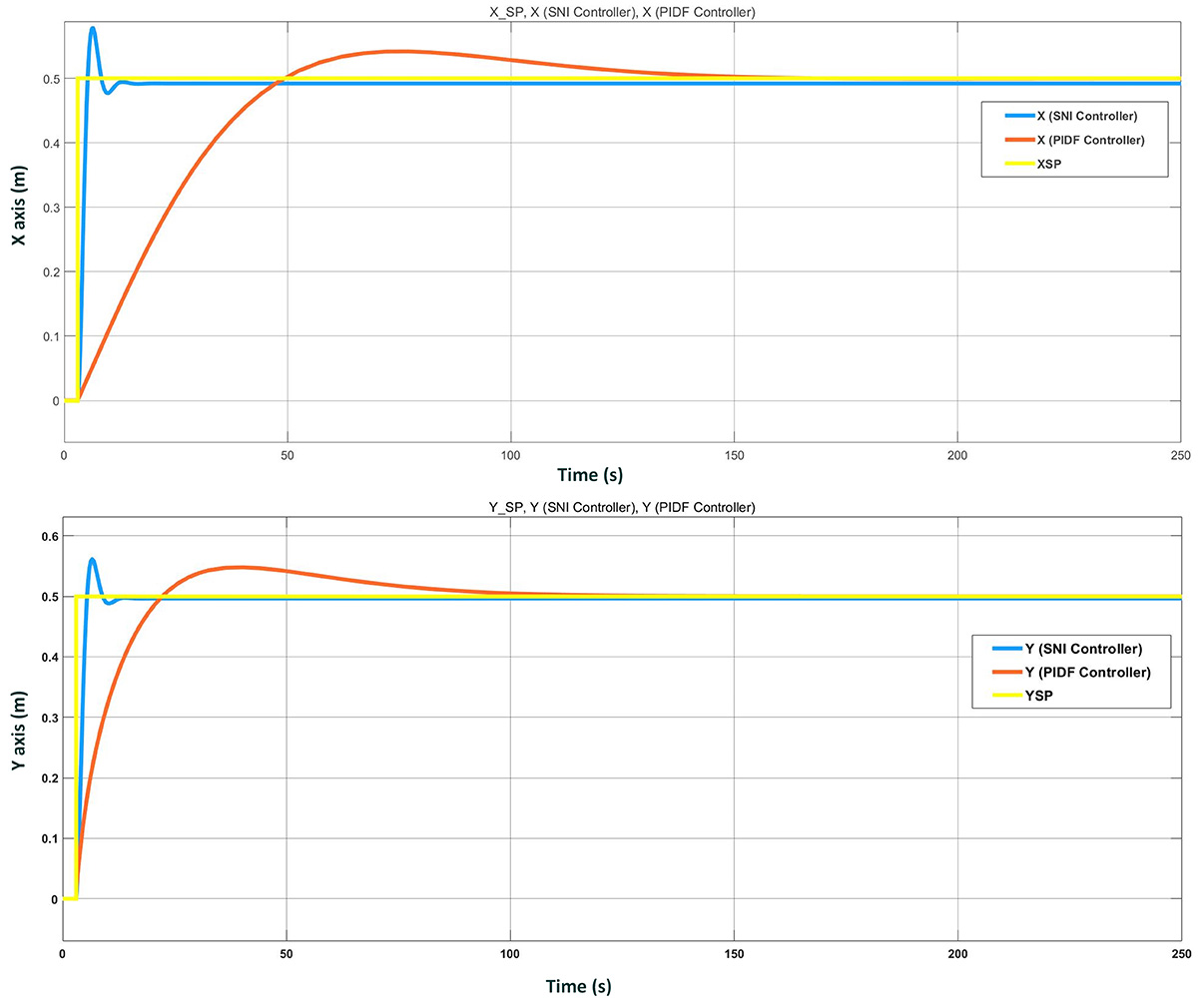}    
		\caption{Performance comparison of the SNI and PIDF controller for vertical and horizontal loop with respect to yellow-colored unit step reference signals. While the blue curves indicate the actual x/y position for SNI controller, the red lines highlight the actual x/y position for PIDF controller.}  
		\label{fig:SNI_Controller}                                 
	\end{center}                                 
\end{figure} 

As shown in Fig. \ref{fig:SNI_Controller}, the percentage of overshoot (PO) referring to its steady-state position on the x and y axis for SNI controller and PIDF controller is:
\begin{equation}
POx_{SNI} = \frac{0.58 - 0.5}{0.5} 100 \% = 16 \%
\end{equation}
\begin{equation}
POy_{SNI} = \frac{0.56 - 0.5}{0.5} 100 \% = 12 \%
\end{equation}
\begin{equation}
POx_{PIDF} = \frac{0.55 - 0.5}{0.5} 100 \% = 10 \%
\end{equation}
\begin{equation}
POy_{PIDF} = \frac{0.55 - 0.5}{0.5} 100 \% = 10 \%
\end{equation}

It is pointed out that the PO values obtained by SNI controllers are higher than those processed by PIDF controllers at only 6 $\%$. However, under the control of SNI controller, the setting time for a UAV  to exceed the final reference value on the x and y axis is approximately 13 seconds while that following the control of PIDF controllers is about 150 seconds.

\section{Real-Time Experimental Tests}

To highlight the efficacy of our approaches, we have conducted indoor experiments using an ARDrone aerial vehicle and three UGVs. Our indoor flight test facility is comprised of 19 VICON Motion Capture Cameras mounted on a rigid frame. A box fan with a maximum airflow of 5 m$/s$ was used as a disturbance generator for all of the experiments. It is placed at a distance of 1.5 meters blowing in the diagonal direction of the positive x and y axes.

\subsection{Mutual Collision Avoidance Problem between UGVs}
\begin{figure}
	\begin{center}
		\includegraphics[width=0.81\linewidth]{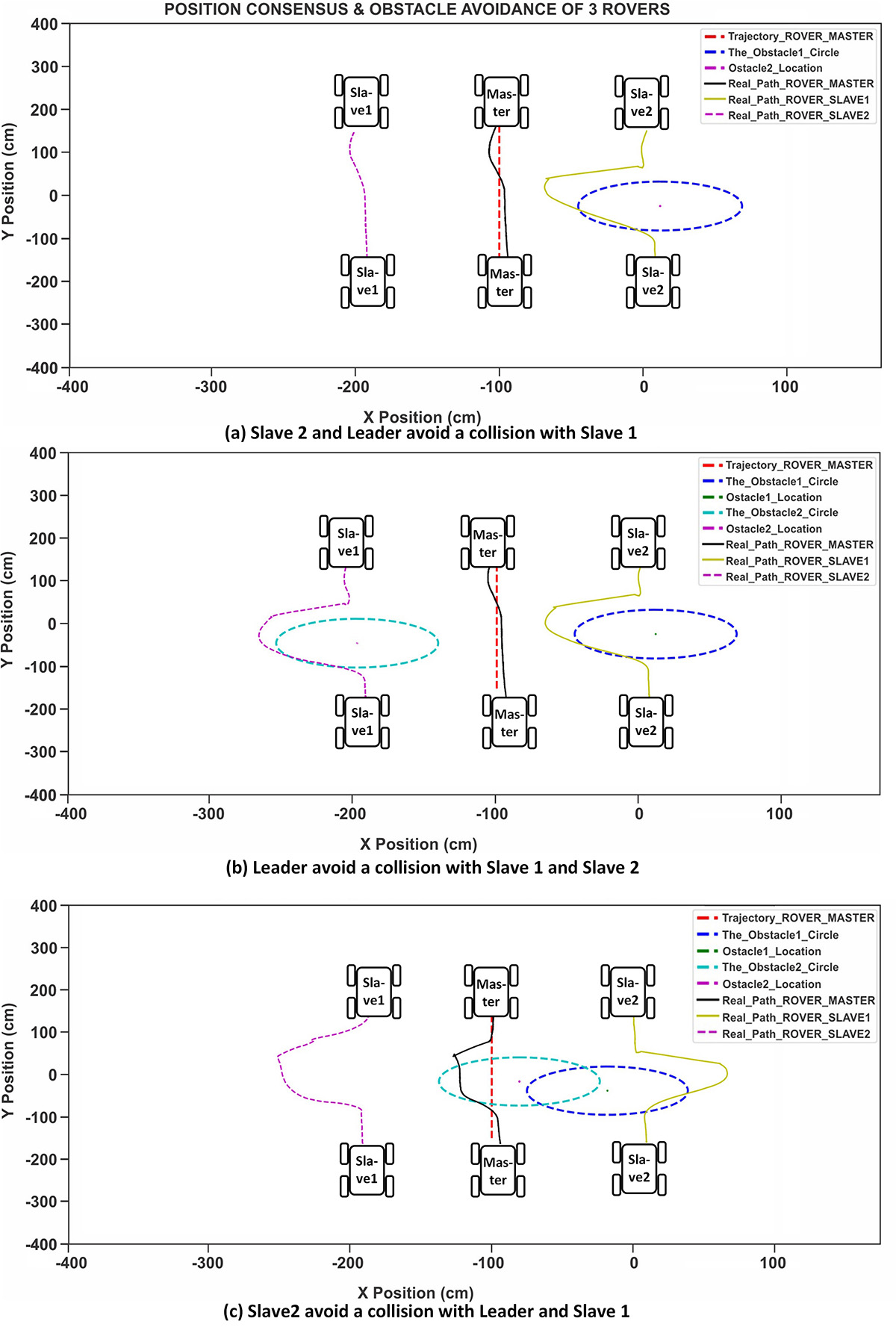}    
		\caption{The implementation of distributed NI inter-vehicle collision avoidance approach in three various scenarios.}  
		\label{fig:mutual_collision}                                 
	\end{center}                                 
\end{figure}

From Fig. \ref{fig:mutual_collision}, it is apparent that our distributed solution can achieve good performance as all three UGVs can avoid mutual collision in any case and safely travel to their target. The fundamental parameters are given as follows: \textit{m} = 12 kg, \textit{K$_{r}$} = \textit{K$_{c}$} = -0.0028, \textit{k$_{r}$} = -0.225, \textit{Vmax$_{sp}$} = 12 cm/s, \textit{r$_{1}$} = \textit{r$_{2}$} = 90 cm, \textit{r$_{ob}$} = 35 cm, target = [-100, 170] cm, dis$_{LF1}$ = [100,0], dis$_{LF2}$ = [-100,0].
 
\subsection{Wind Resistance for a UAV}
Two scenarios have been implemented to test the response of a real quadrotor UAV platform: first, a circular and rectangular trajectory tracking test, with and without added wind gust disturbances; and second, a simple hover test under unknown external forces.

After being tuned in real-time experiments, the SNI-PID and PI-PID controllers have their general forms as below:

\begin{equation}
\begin{split}
M\textunderscore x_{SNI}(s)= M\textunderscore y_{SNI}(s)&= \frac{velx\textunderscore sp}{e\textunderscore posx}=\frac{vely\textunderscore sp}{e\textunderscore posy}\\
&=\frac{-0.35295}{s+1}
\end{split}
\end{equation}
\begin{equation}
\begin{split}
M\textunderscore x_{PI}(s)= M\textunderscore y_{PI}(s)&= \frac{velx\textunderscore sp}{e\textunderscore posx}=\frac{vely\textunderscore sp}{e\textunderscore posy}\\
&= -\frac{0.1374s+0.0021}{s}
\end{split}
\end{equation}

\begin{equation}
\begin{split}
P\textunderscore x(s)&= P\textunderscore y(s)= \frac{roll\textunderscore sp/pitch\textunderscore sp}{e\textunderscore xvel}= \\ &=-\frac{0.3172s+0.0021+0.138s^{2}}{s}
\end{split}
\end{equation}

\subsubsection{Trajectory Tracking Results with The Absence and Presence of Wind Gust Disturbance}
The major purpose of this initial test is to show the SNI-PID controller response and its ability to track circular and rectangular trajectories without being affected by gust disturbance, compared to that of a conventional PI-PID controller. The parameters of circular motion planners are as follows: \textit{r$_{c}$} = 0.8 m, \textit{posx$_{sp}$} = -1.4 + r$_{c}$ $\times$  cos(w$\times$step), \textit{posy$_{sp}$} = -1.2 + r$_{c}$ * sin(w$\times$step);
where \textit{r$_{c}$} is the radius of circular trajectory, step is the time step, \textit{w} = (2$\times$PI/28). Similarly, the parameters of rectangular motion planners are selected as: \textit{posx$_{sp}$} = verx + (0.263$\times$step), \textit{posy$_{sp}$} = very + (0.263$\times$step); where (\textit{verx},\textit{very}) are the four vertexes of a rectangle, including (-1.5, -1.5), (0.6, -1.5), (-0.6, 0.6), and (-1.5, 0.6).

\begin{figure}
	\begin{center}
		\includegraphics[width=0.9\linewidth]{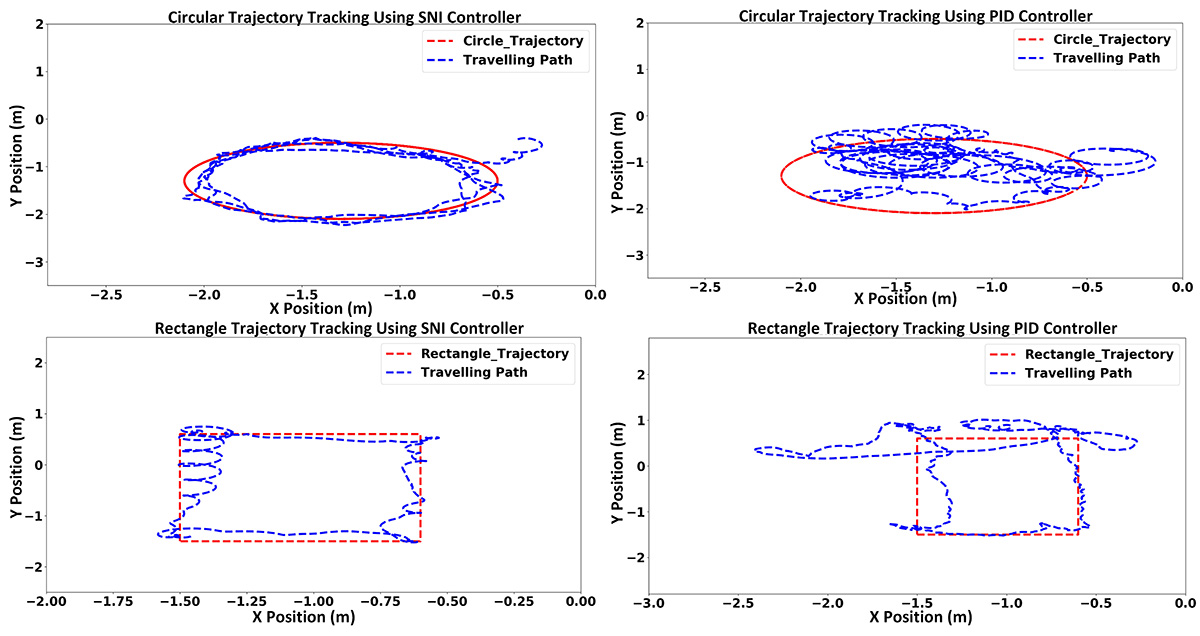}    
		\caption{Performance comparison of our SNI-PID trajectory tracking control system with respect to the conventional PI-PID controller.}  
		\label{fig:crtrajectory_nonoise}                                 
	\end{center}                                 
\end{figure}

\begin{figure}
	\begin{center}
		\includegraphics[width=0.65\linewidth]{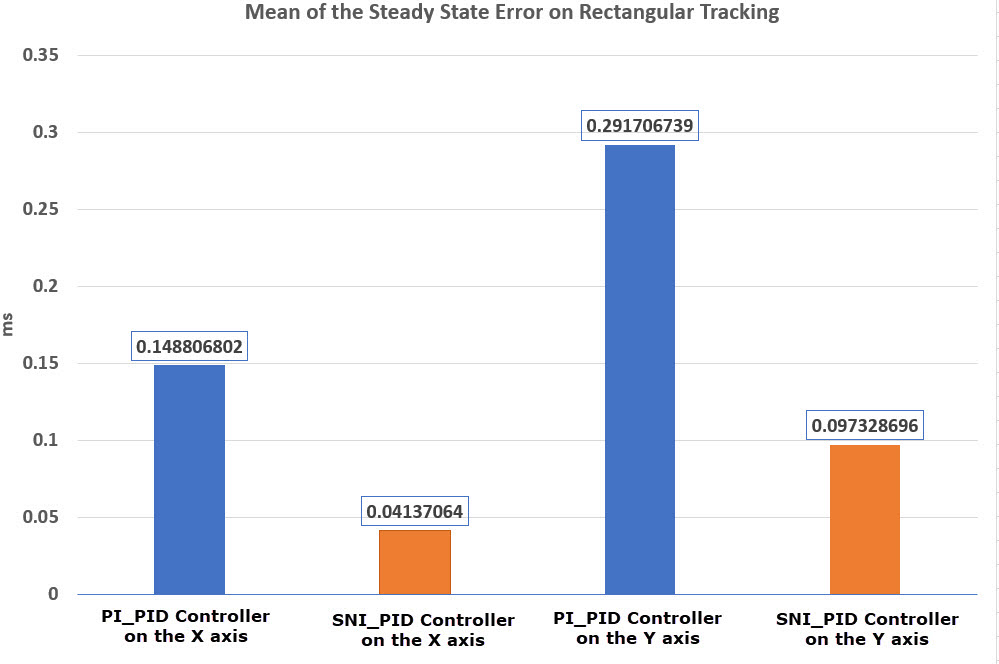}    
		\caption{Statistical measures of the steady state position error performance between our SNI-PID rectangular trajectory tracking controller and PI-PID controller.}  
		\label{fig:rms}                                 
	\end{center}                                 
\end{figure}

Fig. \ref{fig:crtrajectory_nonoise} shows that our SNI-PID controller can easily outperform the conventional PID controllers with better accuracy and smaller steady state error when task completion time is relatively short (28 seconds for drawing a circle and 10 seconds for drawing a side of rectangle). As shown in Fig. \ref{fig:rms}, the Root Mean Square Errors (RMSEs) of SNI-PID controller on the x and y axes are much smaller than those values of PI-PID controller.

\begin{figure}
	\begin{center}
		\includegraphics[width=0.9\linewidth]{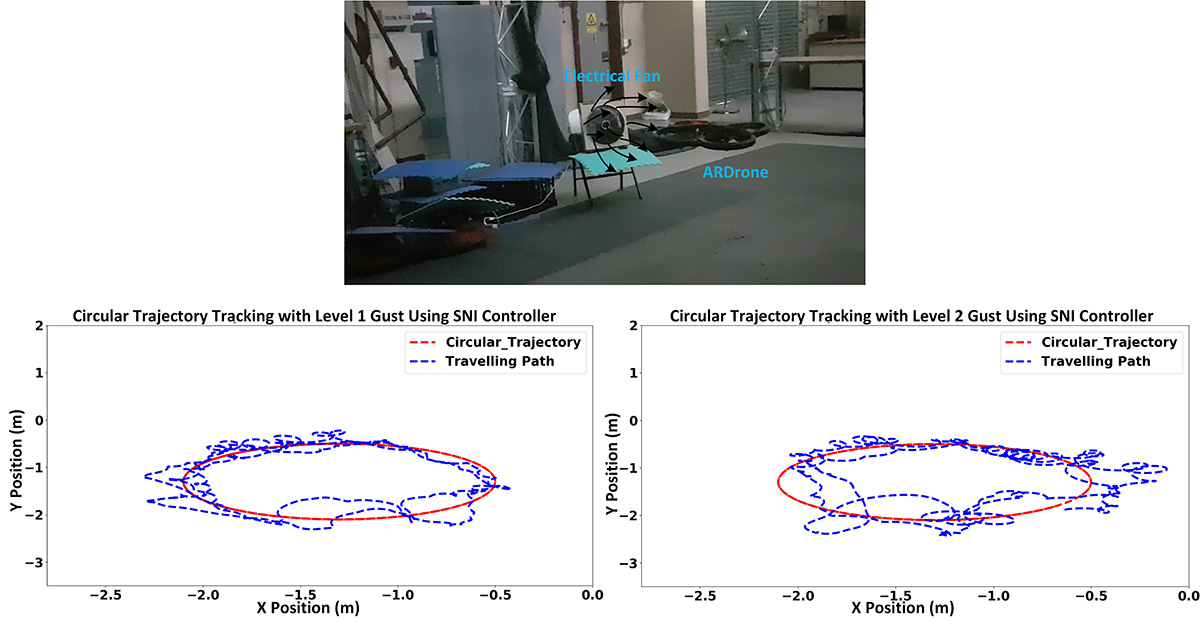}    
		\caption{Position tracking capability of our SNI-PID disturbance-rejection controller with respect to a circle trajectory planner.}  
		\label{fig:ctrajectory_noise}                                 
	\end{center}                                 
\end{figure}

In test 2, our ARDrone suffers two different levels of the wind gust while flying under the guidance of a circular trajectory. Although RMSE of position tracking is larger due to presence of severe disturbances, its performance is acceptable since its actual path still properly follows the track of circular discrete-waypoints for both gust strengths as shown in Fig. \ref{fig:ctrajectory_noise}.

\subsubsection{Hover Results with Wind Gust Disturbance}
\begin{figure}
	\begin{center}
		\includegraphics[width=0.95\linewidth]{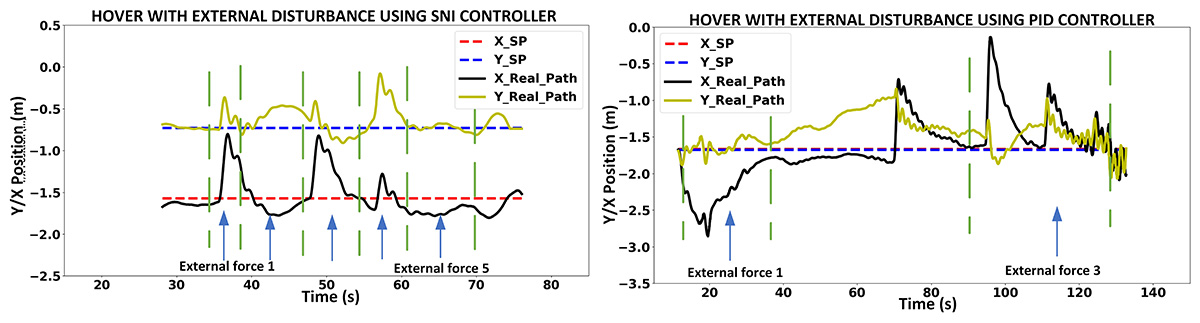}    
		\caption{Hover flight results with unknown external forces.}  
		\label{fig:hover_noise}                                 
	\end{center}                                 
\end{figure}

In Fig. \ref{fig:hover_noise}, the results for hover test while withstanding external forces, whose magnitude and direction are unknown, can be seen. As shown in the figure on the left side, the UAV implementing the SNI-PID position controller easily goes back to the hover point within 6 seconds. In the figure on the right side, under the control of the PI-PID controller, the UAV system takes more time to exceed the hover point again (more than 40 seconds) and occasionally becomes unstable as shown from \textit{t}= 110 seconds to \textit{t}= 135 seconds. It is clear that the SNI-PID controller is able to reject the wind disturbance much faster and with less error than the PI-PID controller.

\subsection{Multiple Obstacles Avoidance Problem for Multi-UGVs with UAV's Assistance}
\begin{figure}
	\begin{center}
		\includegraphics[width=0.86\linewidth]{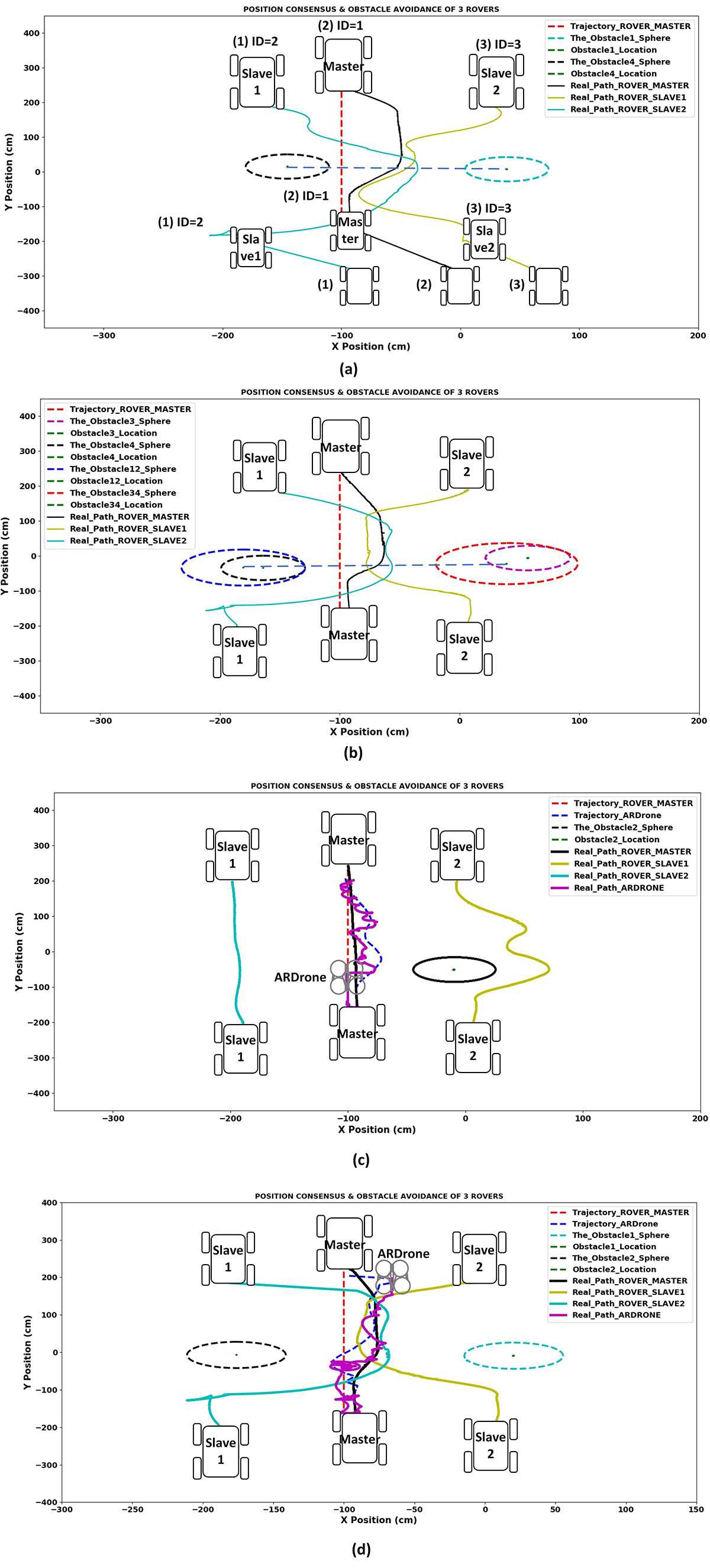}    
		\caption{Real-time performance of the multi-vehicles control system with wind gust disturbances added: (a)(b) Formation and interaction topology variation to avoid multiple obstacles via distributed NI time-varying formation control method and distributed L-F strategy; (c)(d) The air-ground cooperative methodology to solve L/F loss problem in wind gust environment.}  
		\label{fig:queing3}                                 
	\end{center}                                 
\end{figure}

As given in Fig. \ref{fig:queing3} (a) and (b), three UGVs automatically selected their role (L or F) and moved to their initial position in a triangular formation before passing through the narrow space between two arrays of obstacles. Fig. \ref{fig:queing3} (c) and (d) introduce the use of the UAV as a valuable assistance for ground operators by tracking three UGVs simultaneously. It is clear that the maximum position tracking error for all UGVs and UAV is approximately 10 cm on all vertical and horizontal axes although the UGVs' formation is varied continuously and the real flight test is subject to the stronger gust level.   

All videos relative to this literature can be viewed at the address: \url{https://tinyurl.com/y7ygtwac}

\section{Conclusion}
We have demonstrated three new improvements to facilitate the movement of a multi-vehicle system which may face camera constraints, obstacle and inter-robot collision, loss of communications, and gust disturbances.

Compared to the performance of the conventional PID controller, we have shown that the steady-state error of our SNI controller is much lower than that of PID controller in the same scenarios (with and without disturbances). Its superiority is achieved thanks to a faster response and reduced susceptibility to disturbances such as gusts.

With respect to the existing approaches for multi-vehicle formation control, our proposed methods help UGVs to self-arrange and self-decide their role (changing the ID interaction topology) in a pre-defined formation. As a result, UGVs are able to overcome narrow spaces between two arrays of multiple obstacles without any unexpected behaviors or physical constraints as shown in \cite{Dai14, Dai15}. 

Also, since the robot's formation is maintained by the relative distances between L and F as well as between robots and obstacles, instead of being maintained by global robot positions as illustrated in \cite{Dong16}, our methods are also suitable to operate in a limited and unreliable communication environment.

In future work, our SNI gains will be adapted to better cope with multiple disturbances, including both severe external and internal effects. Additionally, a leader-less control method for multiple vehicles will be considered to diminish the adverse impacts of the leader loss situation.



%




\end{document}